\documentclass{jfm}
\usepackage{graphicx}
\usepackage{dcolumn}
\usepackage{bm}
\usepackage{array}
\usepackage{epstopdf}
\usepackage{subfigure}
\usepackage{natbib}
\usepackage{amssymb}
\usepackage{amsbsy}

\title[The role of Stewartson and Ekman layers in RRB convection]
{The role of Stewartson and Ekman layers in turbulent rotating Rayleigh-B\'enard convection}

\author[Rudie P.J. Kunnen et al.]
{Rudie P.J. Kunnen$^1$, Richard J.A.M. Stevens$^2$, Jim Overkamp$^1$, Chao Sun$^2$, GertJan F. van Heijst$^1$, Herman J.H. Clercx$^{1,3}$}

\affiliation{
$^1$Department of Physics and J.M. Burgers Centre for Fluid Dynamics, Eindhoven University of Technology, P.O. Box 513, 5600 MB Eindhoven, The Netherlands\\
$^2$Department of Science and Technology and J.M. Burgers Centre for Fluid Dynamics, University of Twente, P.O. Box 217, 7500 AE Enschede, The Netherlands,\\
$^3$Department of Applied Mathematics, University of Twente, P.O. Box 217, 7500 AE Enschede, The Netherlands}

\begin{document}

\maketitle

\begin{abstract}
When the classical Rayleigh-B\'enard (RB) system is rotated about its vertical 
axis roughly three regimes can be identified. In regime I (weak rotation) the 
large scale circulation (LSC) is the dominant feature of the flow. In regime II 
(moderate rotation) the LSC is replaced by vertically aligned vortices. Regime III 
(strong rotation) is characterized by suppression of the vertical velocity 
fluctuations. Using results from experiments and direct numerical simulations of 
RB convection for a cell with a diameter-to-height aspect ratio equal 
to one at $Ra \sim 10^8-10^9$ ($Pr=4-6$) and $0 \lesssim 1/Ro \lesssim 25$ we identified 
the characteristics of the azimuthal temperature profiles at the sidewall in the 
different regimes. In regime I the azimuthal wall temperature profile shows a cosine 
shape and a vertical temperature gradient due to plumes that travel with the LSC close to the sidewall. In regime II and III this cosine profile disappears, 
but the vertical wall temperature gradient is still observed. It turns out 
that the vertical wall temperature gradient in regimes II and III has a different 
origin than that observed in regime I. It is caused by boundary layer dynamics 
characteristic for rotating flows, which drives a secondary flow that transports 
hot fluid up the sidewall in the lower part of the container and cold fluid downwards 
along the sidewall in the top part. 
\end{abstract}

\section{Introduction}
The classical system to study buoyancy driven flows is the Rayleigh-B\'enard (RB) 
system~\citep{ahl09,loh10}. In this system a layer of fluid is confined between two 
horizontal plates and is heated from below and cooled from above. The RB system is also 
very suitable to study the influence of rotation on heat transport mechanisms. In order 
to do this the RB system is rotated at an angular speed $\Omega$ about its vertical axis. 
Studies about the influence of rotation on heat transport are relevant to understand 
many geophysical and astrophysical flow phenomena such as the global thermohaline 
circulation, convection in the interior of gaseous planets, and convection in the 
outer layer of the Sun. Furthermore, the studies are also very relevant for optimization 
of industrial applications. Therefore, turbulent rotating convection has been studied 
extensively in laboratory experiments~\citep[for example,][]{ros69,bou86,fer91,zho93,liu97,sak97,har99,vor02,kun08e,liu09,nie10,zho10c}, 
direct numerical simulations~\citep[for example,][]{jul96,jul96b,kun06,sch09,kun09,sch10,ste10a,ste10b}, and with combined 
experimental and numerical investigations~\citep{kun08b,zho09b,ste09,kun10,kun10b,wei10}.

In this paper we indicate the rotation rate around the vertical axis by the Rossby 
number $Ro$, which compares the inertial and Coriolis forces in the system. Here, $Ro$ 
is defined as
\begin{equation}
    Ro= \sqrt{\frac{Ra}{PrTa}} = \frac{1}{2 \Omega} \sqrt{\frac{\beta g \Delta}{L}},
\end{equation}
where $Pr=\nu/\kappa$ is the Prandtl number, $Ta=(2\Omega L^2/\nu)^2$ is the Taylor 
number and $Ra=\beta g L^3 \Delta/(\nu \kappa)$ is the Rayleigh number, with $\Omega$ 
the rotation rate, $\beta$ the thermal expansion coefficient, $g$ the gravitational 
acceleration, $\nu$ the kinematic viscosity, $\kappa$ the thermal diffusion coefficient, 
and $\Delta$ the temperature difference between the two plates. Throughout the paper we 
will consider a cell with aspect ratio $\Gamma \equiv D/L=1$, with $L$ the height of the 
RB cell and $D$ its diameter. The dimensionless heat transport in the system is indicated 
by the Nusselt number,
\begin{equation}
    Nu= \frac{QL}{\lambda \Delta}~,
\label{eq:nusselt}
\end{equation}
where $Q$ is the heat-current density and $\lambda$ the thermal conductivity of the fluid in the absence of convection.

\begin{figure}
  \centering
  \includegraphics[width=0.49\textwidth]{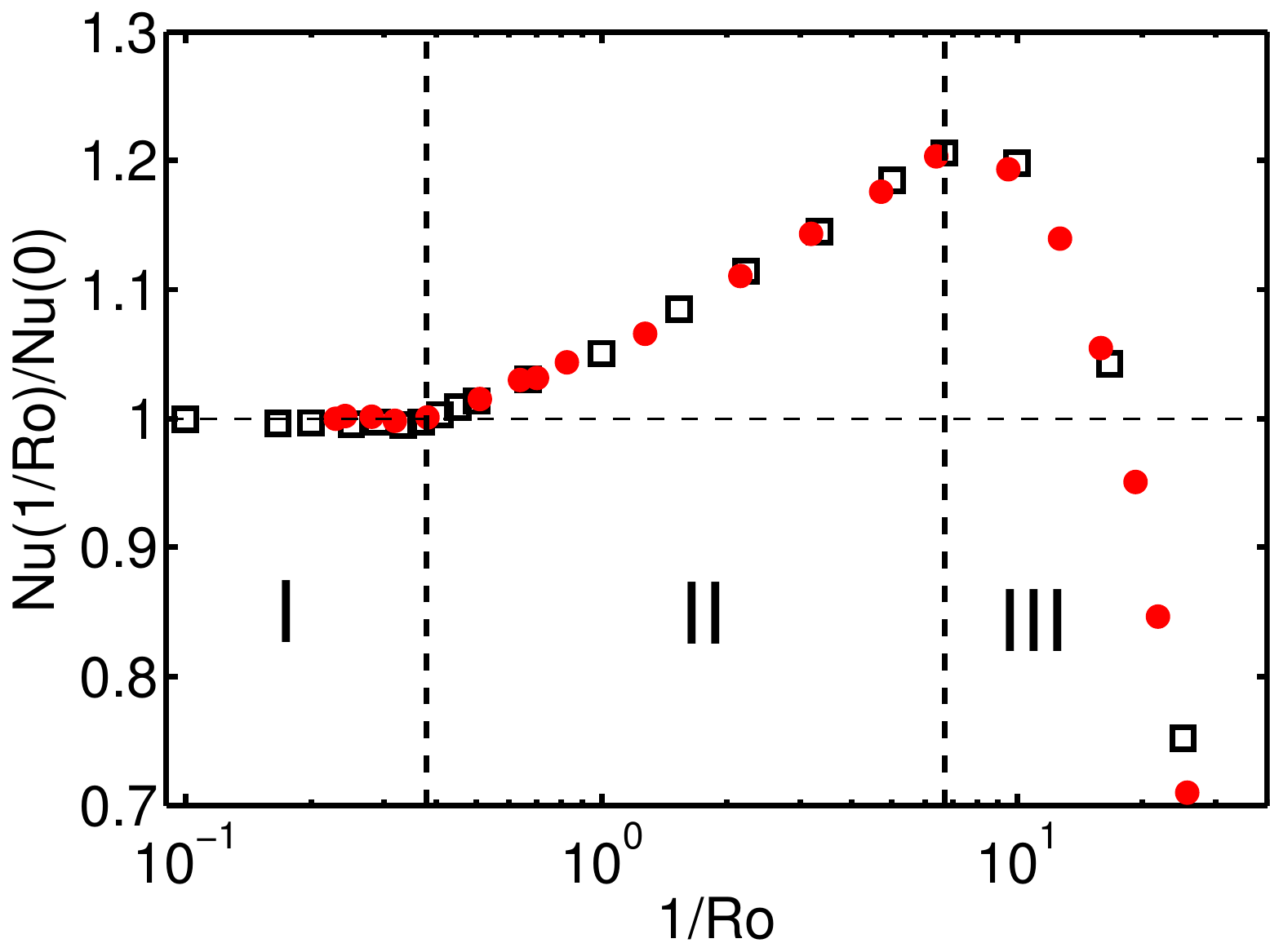}
  \caption{The scaled heat transfer $Nu(1/Ro)/Nu (0)$, with $Nu$ the Nusselt number 
defined in Eq.~(\ref{eq:nusselt}),  as function of $1/Ro$ on a logarithmic scale. Experimental and 
numerical data for $Ra=2.73 \times 10^8$ and $Pr=6.26$ are indicated by red dots and open 
squares, respectively. Data taken from \citet{zho09b} and \citet{ste09} and are called run E2 in \citet{zho10c}. The transition 
between the different regimes, see text, is indicated by the vertical dashed lines.}
  \label{Figure Nusselt}
\end{figure}

In figure \ref{Figure Nusselt} we show a typical measurement of the heat transport 
enhancement with respect to the non-rotating case as function of the rotation rate~\citep{zho09b,ste09}. The figure shows that, depending on the rotation rate, three 
different regimes can be identified. Regime I (weak rotation), where no heat transport 
enhancement is observed, regime II (moderate rotation), where a strong heat transport 
enhancement is found, and regime III (strong rotation), where the heat transport starts 
to decrease. We note that the division between regime I and regime II is obvious as there is a bifurcation~\citep{zho09b,ste09,wei10}. Furthermore, we note that the log plot in figure \ref{Figure Nusselt} makes the transition from regime II to regime III look more sudden than it actually is. \citet{zho09b} and~\citet{ste09} have shown that the position of the onset does not depend on~$Ra$ and~$Pr$. However, the position of the maximum heat transfer enhancement shifts towards lower~$1/Ro$ when~$Ra$ is increased or~$Pr$ is decreased. In addition, the maximum heat transfer enhancement decreases with increasing~$Ra$ and decreasing~$Pr$. For a detailed discussion on the influence of~$Ra$ and~$Pr$ we refer the reader to~\citet{zho09b} and~\citet{ste10a}. Flow visualization experiments~\citep{bou90,kun10} and the analysis of the flow structures obtained in numerical simulations~\citep{ste09,kun10} have confirmed 
that this division of regimes coincides with changes observed in the flow patterns and 
flow characteristics.

In regime I ($1/Ro\lesssim0.5$) the large scale circulation (LSC), typical for 
non-rotating RB convection, is still present, because the Coriolis force is too weak to 
overcome the buoyancy force that causes the LSC. \citet{ste09} showed that there is a 
sharp transition to the regime where rotational effects become important, while at the 
same time the strength of the LSC is decreasing~\citep{kun08b}. 
\citet{zho10c} experimentally found that the time-averaged LSC amplitudes decrease strongly at the transition from regime I to regime II, see figure 13 of their paper.
In regime II ($0.5 \lesssim 1/Ro \lesssim 6.67$) the LSC is replaced by predominantly vertically 
oriented vortical columns as the dominant flow structures and a large increase in the 
heat transport is observed.

This enhanced heat transport has been ascribed to Ekman 
pumping~\citep{ros69,jul96b,vor02,kun08b,kin09,ste09}. The effect of Ekman pumping, 
and thus the observed heat transfer enhancement, depends strongly on the $Ra$- and $Pr$-values~\citep{zho09b,ste10a}. When the rotation rate is increased further a large 
decrease in the heat transport is observed, because the vertical velocity fluctuations 
are suppressed due to the rotation~\citep{kun08e,kun10}. We will call this regime III.

In this paper we address the question of how these different regimes can be identified from measurements of the azimuthal wall temperature distribution and the vertical temperature gradient along the sidewall, with probes that are embedded in the sidewall of a RB convection cell. This method of sidewall temperature measurements has been introduced by~\citet{bro05b} and a validation of this method is described in the second paragraph of Section 2 of~\citet{bro06}. Recently, \citet{zho10c} have extensively studied the properties of the LSC in RRB convection in aspect ratio $\Gamma=1$ experiments using this method. These measurements covered the $Ra$ number range $3 \times 10^8 \lesssim Ra \lesssim 2 \times 10^{10}$, the $Pr$ number range $3.0  \lesssim Pr \lesssim 6.4$, and the $1/Ro$ number range $0 \lesssim 1/Ro \lesssim 20$. In these measurements detailed statistics about the thermal LSC amplitude (i.e. the amplitude of the cosine fit to the azimuthal temperature profile at the sidewall), the LSC orientation over time, the temperature gradient along the sidewall, the retrograde rotation of the LSC, the frequency of cessations, etc., over this wide parameter range were determined.

This paper is organized as follows. The experimental setup that has been built at the Fluid Dynamics Laboratory at Eindhoven University of Technology is discussed in section 2. The design of this setup is closely based on the Santa Barbara design~\citep{bro05}. Subsequently we will present some experimental results on the properties of the azimuthal temperature profiles in section 3. For a much more detailed set of experimental results on the properties of the azimuthal temperature profiles we refer to the paper of~\citet{zho10c}. In Section 4 we compare the experimental data with the azimuthal temperature and vertical-velocity profiles close to the sidewall found in direct numerical simulations (DNS). In order to explain the sidewall temperature measurements we study the azimuthally averaged flow profiles obtained from DNS. The analysis presented in section 5 reveals that the sidewall temperature measurements, particularly the presence of the vertical wall temperature gradients in regimes II and III~\citep{zho10c}, can be explained by the presence of Stewartson layers that are formed along the sidewall.

\section{Experimental setup}

\begin{figure}
  \centering
  \includegraphics[width=0.49\textwidth]{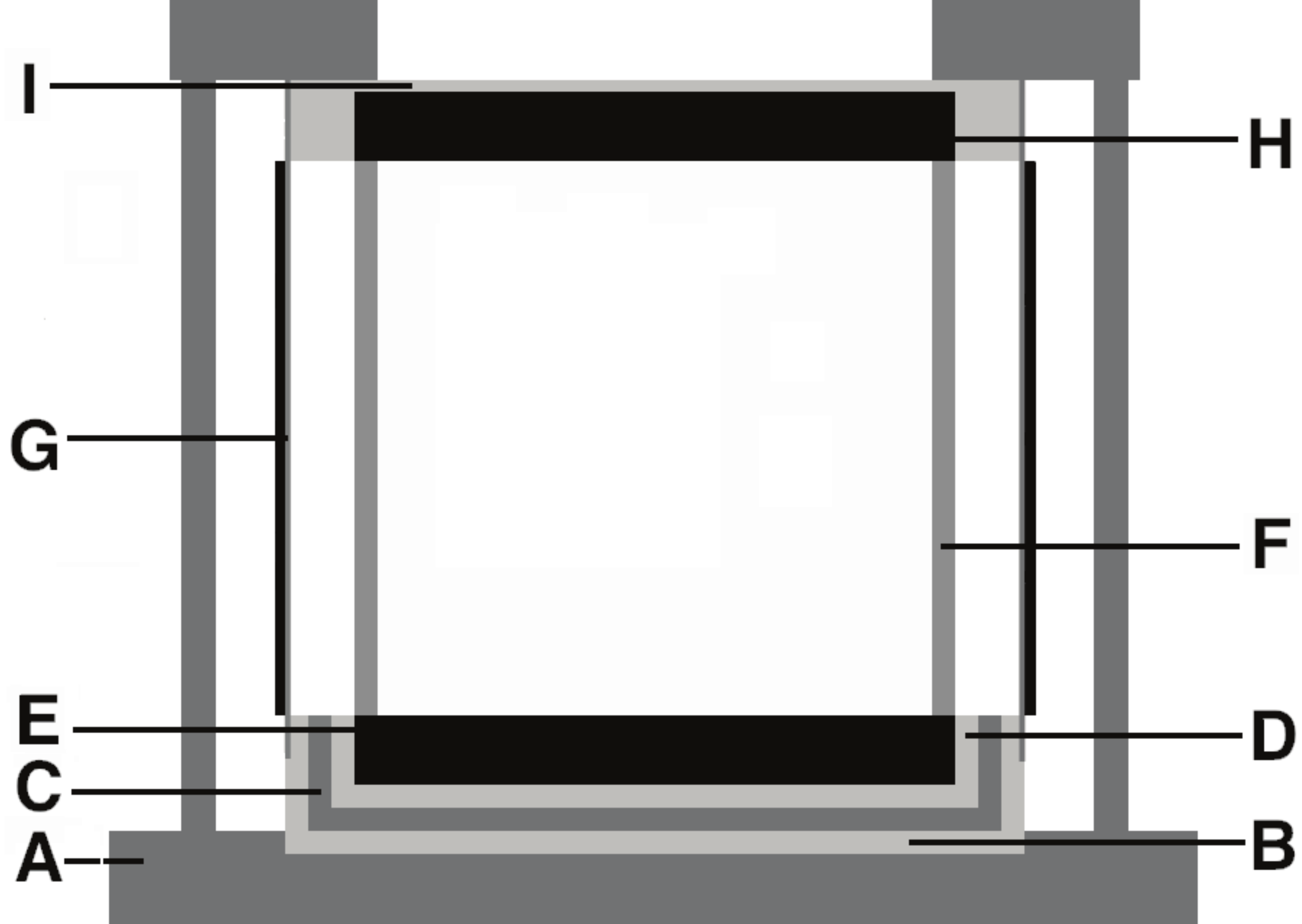}
  \caption{Schematic diagram (not to scale) of the Eindhoven RB apparatus. From bottom 
to top the figure shows the support plate (A), an insulation layer (B), the bottom 
adiabatic shield (C), an insulation layer (D), the bottom copper plate (E), the Plexiglas 
sidewall (F), the adiabatic side shield (G), the top copper plate (H), and the Plexiglas 
top plate (I). See further details in the text.}
  \label{Figure RBsetupEindhoven}
\end{figure}

\noindent Based on the setup described by \citet{bro05} we have built a new RB setup in 
Eindhoven, which is suitable for high precision heat transport measurements. The 
convection cell has a diameter $D$ and height $L$ of 250 mm, making the aspect ratio 
$\Gamma=D/L=1.0$. The modular design of the setup provides the ability to perform 
measurements at different aspect ratios as well, although this paper will only describe 
the $\Gamma=1$ setup in detail. During rotating experiments the RB cell is placed on the 
Eindhoven Rotating Table Facility (RTF)~\citep[for details, see][]{bok07}. A rotating 
connection is available for coolant fluid, with separated in and out flow tubes. The 
rotation rate $\Omega$ of the RTF can be controlled from 0 to 10 rad/s with a resolution 
of 0.001 rad/s and a relative accuracy of $0.5\%$. The rotation rates were kept below 
$\Omega=1.57$ rad/s, yielding Froude numbers $Fr = \Omega^2(D/2)/g$ less than $0.03$ for 
all runs and much smaller for most. This means that the effect of centrifugal forces can 
be considered negligible.

A schematic diagram of the RB cell is shown in figure \ref{Figure RBsetupEindhoven}. From 
bottom to top, we first find the support plate $A$ (400 x 400 mm aluminium), used to 
mount the setup on the RTF and to align the rotational axes of the table and the 
convection cell so that the offset of the cylinder axis and the rotation axis is less 
than 0.1 mm.  Next, part $B$ is a 10 mm insulation layer to prevent heat loss of the 
bottom adiabatic shield $C$ (310 mm outer diameter, 10 mm thick aluminium) to the support 
plate. The shield is fitted with a $250$ W heater to actively control its temperature. 
Inside this adiabatic shield, part $D$ is another 10 mm insulation layer, followed by 
the copper bottom plate $E$ (270 mm outer diameter, 30 mm thick) of the convection cell. 
The back of the bottom plate is covered uniformly with two double spiral grooves of 2 mm 
depth, 2 mm width and 6 mm spacing. Two 4 m, 12.3 $\Omega$ resistance wires are epoxied 
into these grooves. From the back of the plate, five small holes (one at the centre, two 
at a radius of 98 mm and two at 100 mm, i.e. in the space that is available between the 
heater wires that are placed inside the plate) are drilled to within 0.7 mm of its top 
surface and thermistors are mounted in these holes.

On top of the bottom plate, the Plexiglas sidewall $F$ with an inner diameter of 250 mm 
is placed. The thickness of the Plexiglas sidewall is 10 mm to provide enough strength 
in strongly RRB experiments. A rubber O-ring seals the interface between the 
bottom plate and the sidewall and prevents any leakage of fluid. The sidewall contains 
a fluid inlet near the bottom plate and, at opposite angular position, an outlet adjacent 
to the top plate. Care has been taken to avoid the entrapment of air in the convection 
cell, both when filling the cell and when heating the fluid. The sidewall is surrounded 
by an adiabatic side shield $G$ made of two 3 mm thick aluminium plates. The shield is 
actively temperature controlled, ensuring that its temperature is always close to the 
mean temperature of the system. Additional insulation is present between the Plexiglas 
side wall and the aluminium side shield.

The top of the convection cell is formed by the copper top plate $H$, which has similar 
dimensions as the bottom plate. The top plate contains a double spiral water channel of 
8 mm width, 26 mm depth and 25 mm spacing. The water, coming from a refrigerated 
circulator (Thermo Scientific HAAKE DC50-K41) at 12.5 L/min, cools the top plate 
down to the desired temperature. From the top of the plate, five small holes (one at 
the centre, four at a radius of 100 mm) were drilled to within 0.7 mm of the copper-fluid 
interface and thermistors are mounted in these holes. The top plate is covered by a 
Plexiglas top $I$ and the water channel is sealed with a rubber ring. The visual access 
provides a way to check the proper operation of the cooling system. An aluminium 
construction ring is connected to the upper side of the top plate. It is supported by 
6 stainless steel support poles. With these poles, the RB cell can be fixed to achieve 
a fully watertight connection between the plates and the sidewall. To decrease the 
effects of air convection near the convection cell, the entire construction is placed 
in a wooden box covered with a 40 mm insulating layer on the inside.

The apparatus contains 40 thermistors, which were calibrated simultaneously in a separate 
apparatus against a laboratory standard. During the calibration procedure all thermistors 
and the laboratory standard are placed in an extremely well temperature-controlled batch 
(temperature differences less than 0.002 K) to determine the resistance at a set of known 
temperatures. From this calibration data fits were made with 5-th order logarithmic 
polynomials to calculate temperature values from 3-wire resistance measurements. 
Deviations from these fits are generally less than 0.001 K. During the calibration 
procedure all thermistors are connected to the same software and hardware as in the real 
experiments. 24 of these thermistors are placed in the sidewall, forming 3 rings of 8 
equally spaced sensors at heights $0.25L$, $0.5L$ and $0.75L$. Both the top and bottom 
plate contain 5 sensors to monitor the temperature of the plate. The remaining sensors 
are used to control the temperature of the bottom adiabatic shield and of the insulation 
around the Plexiglas sidewall. Readings of all thermometer resistances and of the bottom 
plate heater current and heater voltage were taken every second. The top (bottom) 
temperature $T_t$ ($T_b$) was set equal to the area average of the five thermometers 
embedded in the top (bottom) plate. For any given data point, measurements over typically 
the first four hours were discarded to avoid transients, and data taken over an 
additional period of at least another eight hours were averaged to get the heat-current 
density $Q$, and the temperatures $T_b$ and $T_t$. Because of the imperfect temperature uniformity of the bottom shield there was a small 
parasitic heat loss from the bottom plate of about $0.1$ W, which was determined by 
measuring the required power to keep the bottom and top plate at $40 ^\circ C$, a 
measurement which takes two days. For each measurement this parasitic heat loss was subtracted from the measured power.

\begin{figure}
  \centering
  \includegraphics[width=0.49\textwidth]{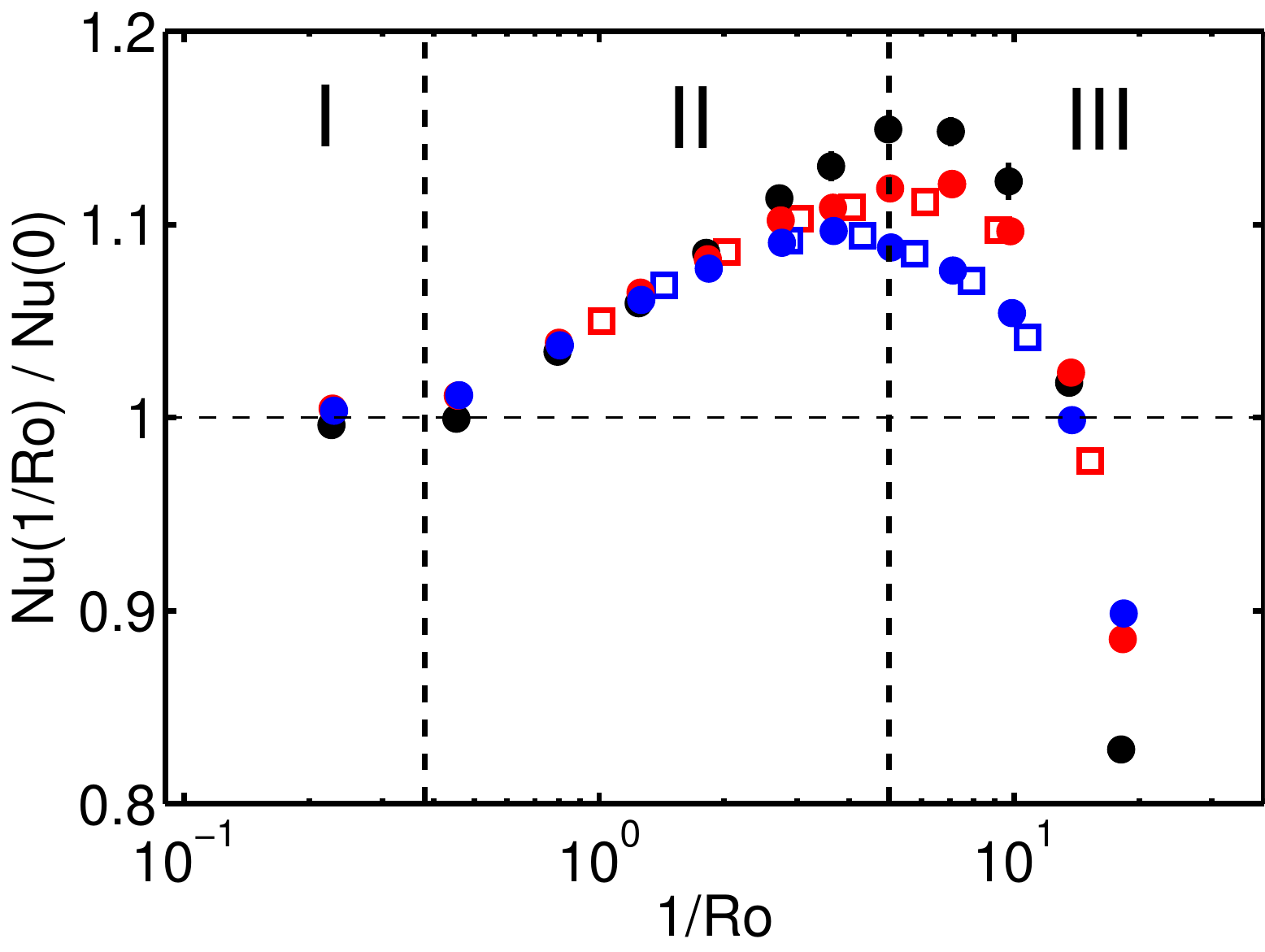}
  \caption{The ratio of the Nusselt number $Nu(1/Ro)$ in the presence of rotation to 
$Nu(0)$ for $Pr=4.38$ ($T_m = 40.00^\circ$C) and $\Gamma=1$. The solid circles are the 
data obtained in the Eindhoven RB setup and the open squares are the corresponding data 
obtained in Santa Barbara~\citep{zho09b,ste09}. Black solid circles: 
$Ra = 2.99\times 10^8$ ($\Delta = 0.50$ K). Red solid circles: $Ra = 5.88\times 10^8$ 
($\Delta = 1.00$ K). Blue solid circles: $Ra = 1.16\times 10^9$  ($\Delta = 2.00$ K). 
Red open squares: $Ra = 5.6\times 10^8$~\citep[$\Delta = 1.00$ K, run E4 in][]{zho10c}. Blue open squares: 
$Ra = 1.2\times 10^9$ \citep[$\Delta = 2.00$ K,  run E5 in][]{zho10c}. }
  \label{Figure NusseltGamma1Eindhoven}
\end{figure}

The measurements have been verified against the experiments of \citet{fun05}, \citet{ahl09} 
and \citet{ste09}. For the present measurements the difference is in the order of $1\%$, 
well within the differences observed between different RB setups~\citep{ahl09}. In 
figure \ref{Figure NusseltGamma1Eindhoven} we show the heat transfer enhancement 
$Nu(\Omega)/Nu(0)$ as function of the rotation rate $1/Ro$ for 
$Ra=2.99\times10^8$, $Ra=5.88\times10^8$ and $Ra=1.16\times10^9$ at $Pr=4.38$, which 
corresponds with a mean temperature of the fluid of $40.00^\circ$C, and compare them 
with similar heat transport measurements reported in the literature~\citep{zho09b,ste09}. We find that the two datasets involving rotation also agree within~$1\%$, which can be considered a very good agreement.

\section{Experimental sidewall measurements} \label{section Experiments}

\noindent In RB experiments it is common to determine the flow properties by analyzing 
the azimuthal wall temperature profile obtained by thermistors that are embedded in the 
sidewall~\citep{ahl09} and in our setup we do the same. Following \citet{ste10c} we define 
the relative LSC strength at midheight ($\bar{S}_m$), based on the energy in the different 
modes of the azimuthal temperature profile, as
\begin{equation}\label{Eq Relative Strength LSC}
    \bar{S}_m = \mathrm{max}\left( \left( \frac{\sum_{t_b}^{t_e} E_1}{\sum_{t_b}^{t_e} E_{tot}} - \frac{1}{N}\right) / \left(1-\frac{1}{N}\right), 0 \right).
\end{equation}
Here 
$\sum_{t_b}^{t_e} E_1$ indicates the sum of the energy in the first Fourier mode over time, i.e. from the beginning of the simulation $t=t_b$ to the end of the simulation $t=t_e$, $\sum_{t_b}^{t_e} E_{tot}$ the sum of the total energy in all Fourier modes over time, and $N$ the total number of Fourier modes that can be determined. 
The relative LSC strength $\bar{S}_m$ always 
has a value between $0$ and $1$; here $1$ indicates that the azimuthal profile is a pure 
cosine profile, which is a signature of the LSC according to \citet{bro06}, and $0$ indicates that the magnitude of 
the cosine mode is equal to (or weaker than) the value expected from a random noise 
signal. Hence $\bar{S}_m \gg 0.5$ indicates that a cosine fit on average is a 
reasonable approximation of the data, as then most energy in the signal resides in the 
first Fourier mode.  In contrast, $\bar{S}_m \ll 0.5$  indicates that most energy 
resides in the higher Fourier modes. Hence, we consider the LSC as dominant once 
$\bar{S}_m \gg 0.5$ at midheight. A small value of $\bar{S}_m$ indicates that 
no single LSC is found, implying the existence of either multiple rolls or, in the case 
of rotating Rayleigh-B\'enard (RRB) convection, vertically aligned vortices.

\begin{figure}
  \centering
  \subfigure[]{\includegraphics[width=0.49\textwidth]{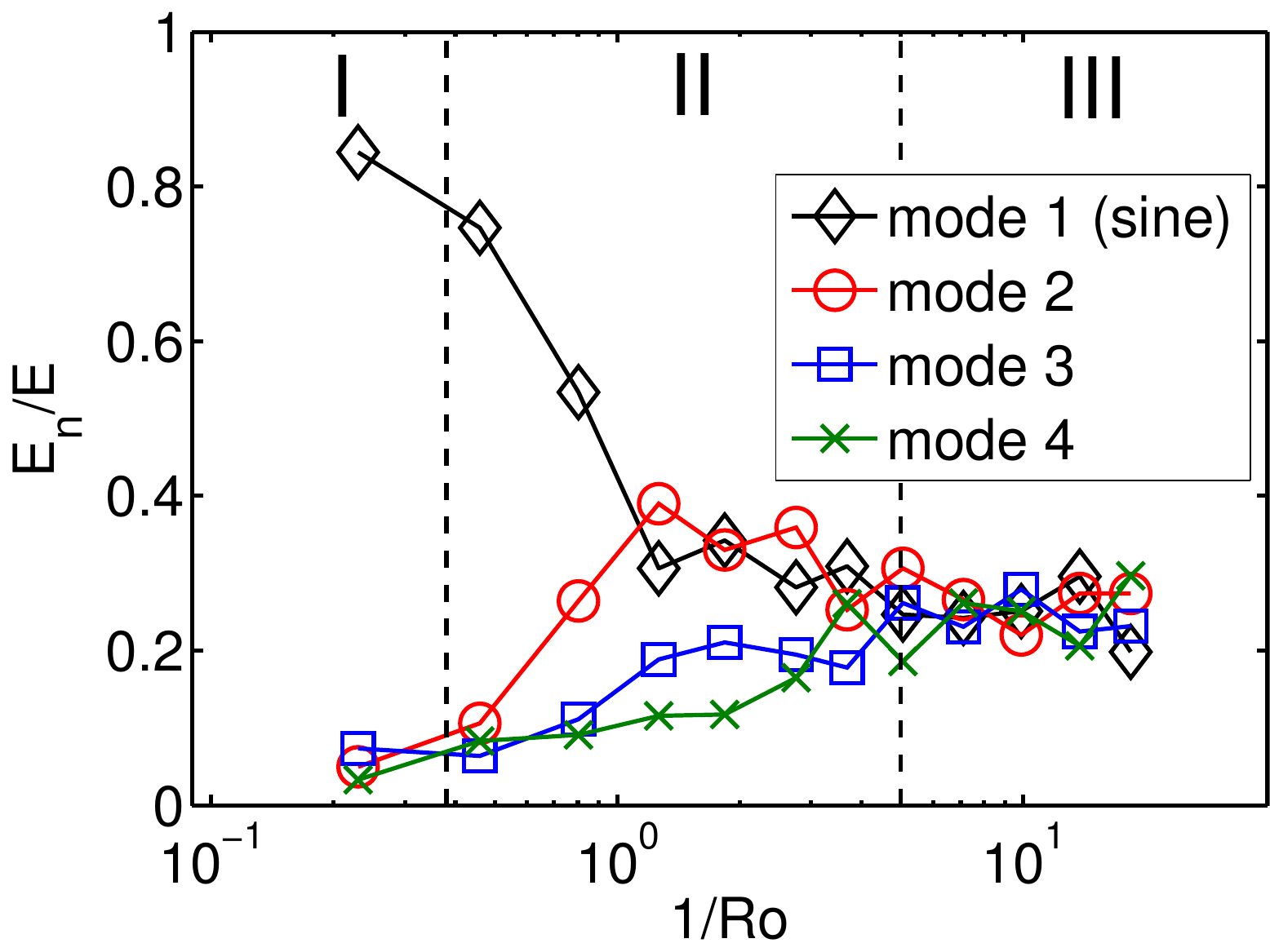}}
  \subfigure[]{\includegraphics[width=0.49\textwidth]{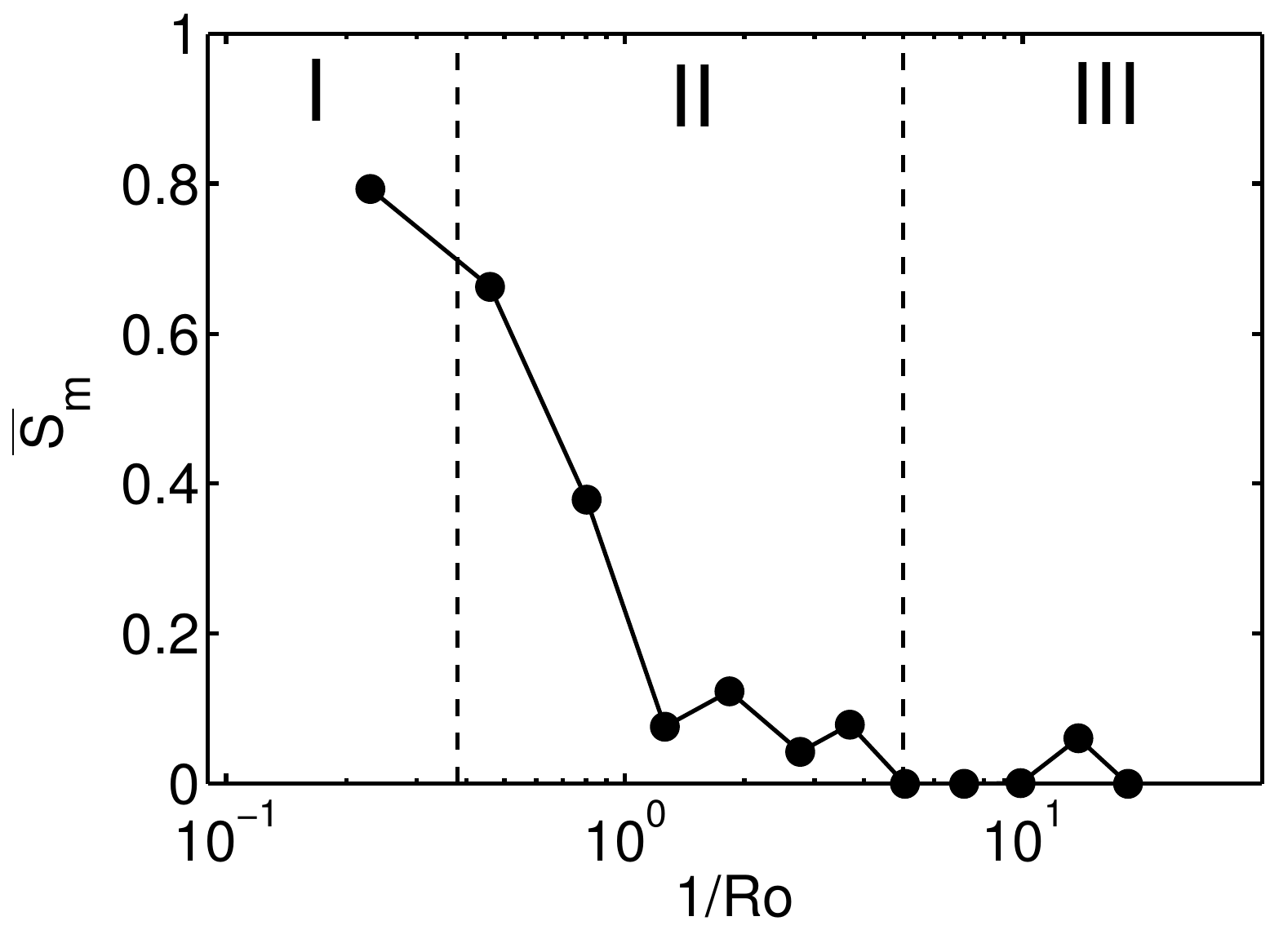}}
  \caption{Experimental results on the magnitude of the different Fourier modes and 
$S_m$ for $Ra=1.16\times10^9$ and $Pr=4.38$ based on the data of $8$ azimuthally 
equally spaced probes at $z=0.50L$. a) The energy in the different Fourier modes of the 
azimuthal wall temperature profile normalized by the energy present in all Fourier modes 
as function of $1/Ro$. b) The corresponding relative LSC strength $S_m$. The two 
dashed vertical lines in both panels indicate the transitions between regimes I and II, 
and between II and III, respectively.}
  \label{Figure Fourier experimental}
\end{figure}

Figure \ref{Figure Fourier experimental}(a) shows the magnitude of the different Fourier 
modes of the azimuthal wall temperature profile for the experiments at 
$Ra=1.16\times10^9$ with $Pr=4.38$ based on the data of eight equally spaced thermistors 
placed inside the sidewall (located at $z=0.5L$). The corresponding relative LSC strength 
is given in panel b. The figure clearly shows that the relative LSC strength is large in 
regime I, which indicates the presence of a LSC. However, for higher rotation rates, 
i.e. regime II, the relative LSC strength decreases because vertically aligned vortices 
become the dominant feature of the flow. For this flow one expects a random azimuthal temperature profile at midheight, which is confirmed by the low relative LSC strength $S_m$ in regime II and III, see figure \ref{Figure Fourier experimental}(b).

Extensive sidewall temperature gradient measurements for non-rotating RB convection were done by \citet{bro07} and for RRB convection by \citet{zho10c}, see their figures 10 and 11. When the LSC is the dominant feature of the flow the vertical temperature gradient at the sidewall is mainly due to plumes that travel close to the sidewall with the LSC. Thus one would expect that the temperature gradient along the sidewall should decrease in regime II as the LSC disappears there. However figure 10 of \citet{zho10c} shows that the temperature gradient along the sidewall even increases in regime II and III. This nonzero temperature gradient along the sidewall in regime II and III  is caused by the secondary flow that will be discussed in section \ref{section_Ekman_Stewartson}. Further interesting information about the LSC can be found in figure~$13$ of~\citet{zho10c}. The figure shows that the temperature amplitude of the LSC, i.e. the strength of the first Fourier mode, starts to decrease at~$1/Ro\approx 0.41$ and that the LSC ceases to contribute significantly for~$1/Ro\gtrsim 0.8$. Thus our experimental results in figure~\ref{Figure Fourier experimental} are in good agreement with the results of~\citet{zho10c}. It is interesting to note that the measurements of~\citet{zho10c} reveal an increase in the LSC strength between~$1/Ro=0$ and~$1/Ro\approx 0.41$. In our data this effect is not visible, since we do not have enough measurement points in regime I. As is discussed by~\citet{zho10c} there is no well-accepted theoretical explanation for this phenomenon at the moment.

\section{Numerical study of temperature and vertical-velocity profiles close to the sidewall}

\noindent We shall now consider the results of DNS of RRB convection performed at various~$Ro$ values for~$Ra=2.73\times 10^8$ and~$Pr=6.26$~\citep[see details in][]{zho09b} and for~$Ra=1.00\times 10^9$ and~$Pr=6.4$~\citep[see details in][]{kun10}, which are in the same parameter regime as the experiments discussed above. In the simulations we solved the three-dimensional Navier-Stokes equations within the Boussinesq approximation in a cylindrical cell with $\Gamma=1$ with no-slip boundary conditions at all walls, a uniform temperature at the horizontal plates, and an adiabatic sidewall. For further details about the numerical 
code we refer to \citet{ver96} and \citet{ver97,ver03}. In all simulations we calculate 
the azimuthal averages of the three velocity components and the temperature. Furthermore, 
in the simulations at $Ra=2.73\times10^8$ we placed $32$ azimuthally equally spaced 
numerical probes that provide simultaneous point-wise measurements of the temperature 
and the three velocity components $u_r$, $u_{\phi}$, $u_z$ in the radial, azimuthal, and 
vertical directions $r$, $\phi$ and $z$, respectively, at the heights $0.25L$, $0.50L$, 
and $0.75L$ and a distances $0.45 L$ from the cylinder axis. The azimuthal temperature 
and vertical-velocity profiles measured by the probes are analyzed in the same way as 
the experimental data.

\begin{figure}
  \centering
  \subfigure[]{\includegraphics[width=0.49\textwidth]{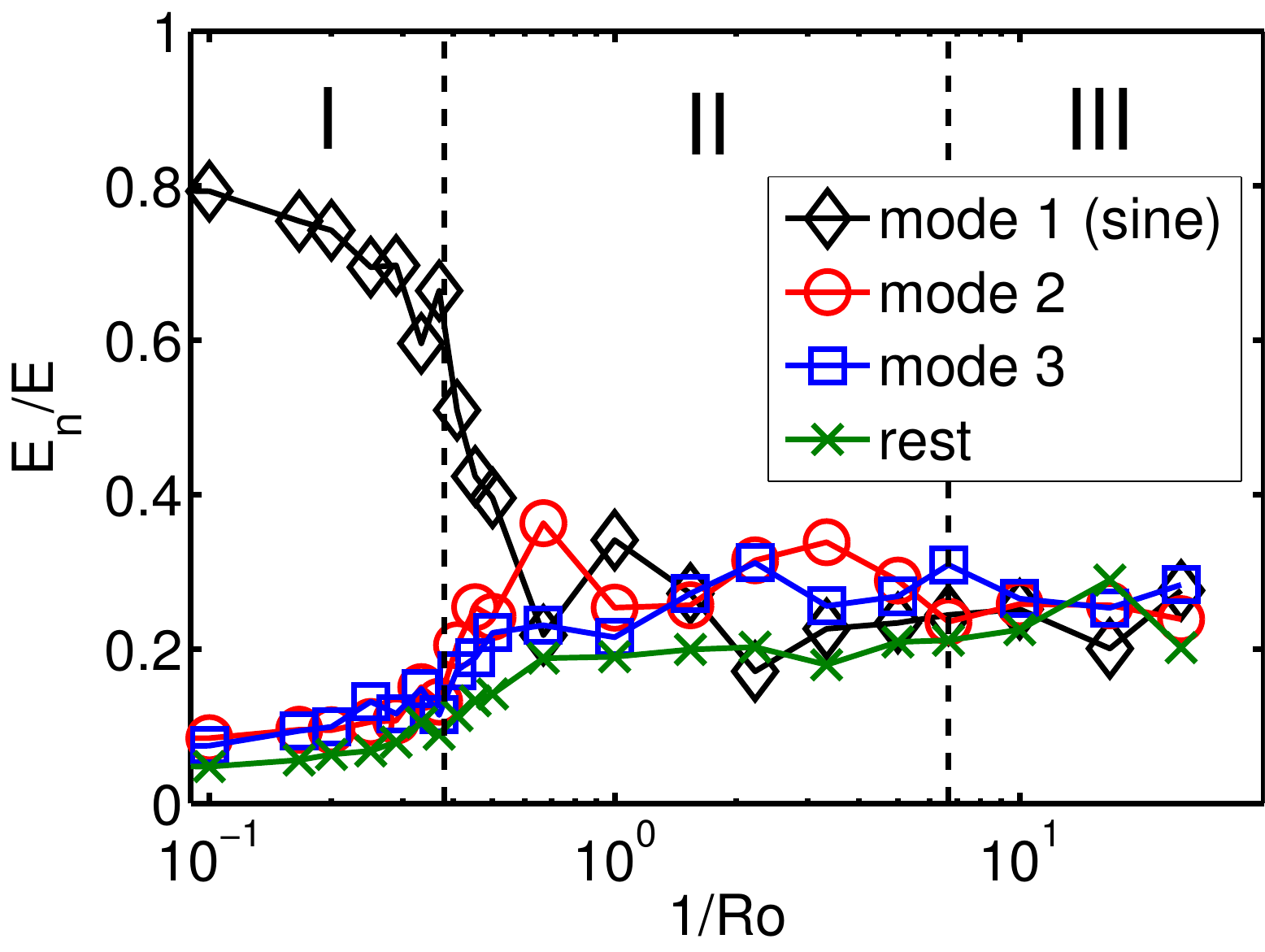}}
  \subfigure[]{\includegraphics[width=0.49\textwidth]{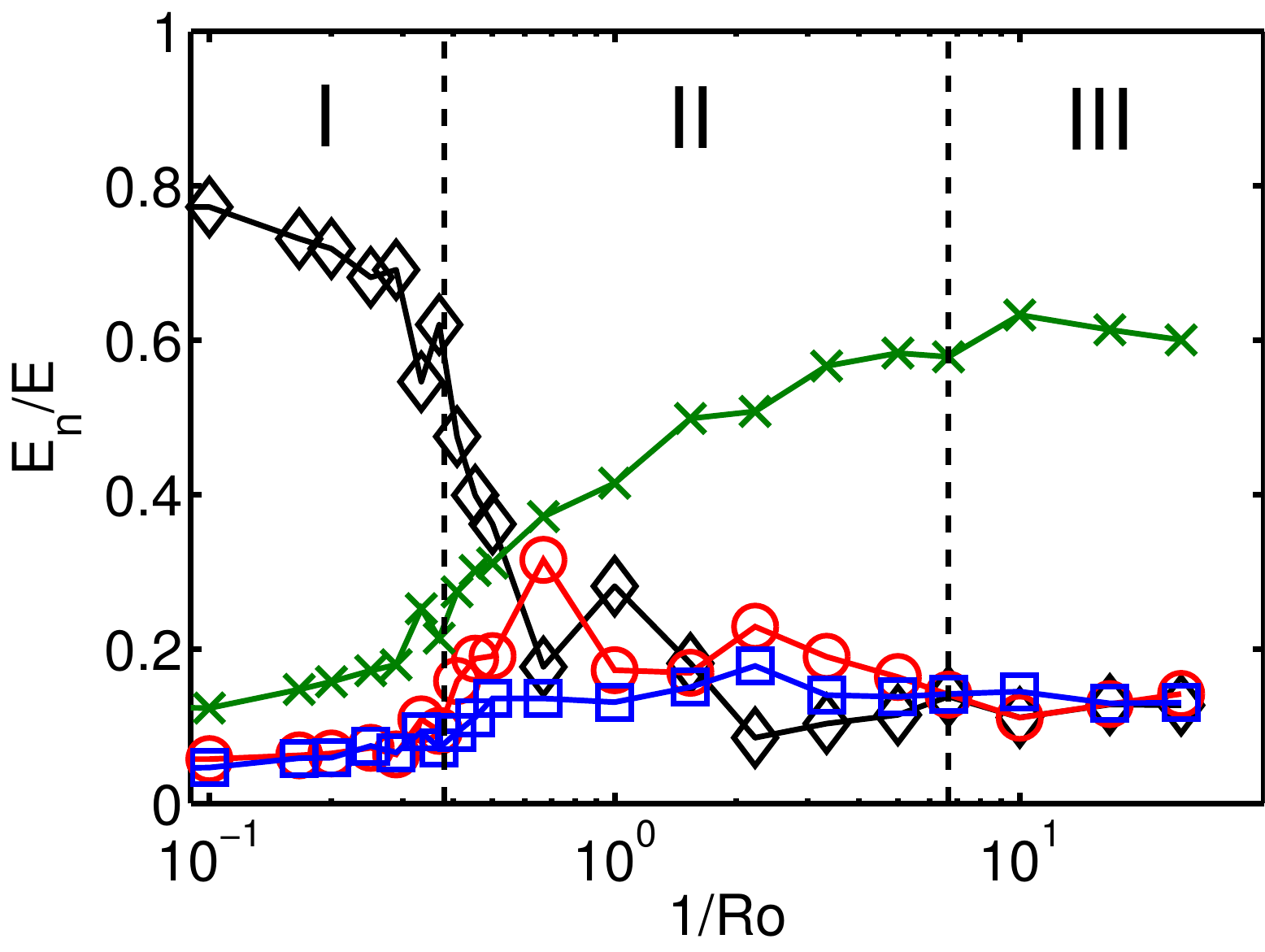}}
  \subfigure[]{\includegraphics[width=0.49\textwidth]{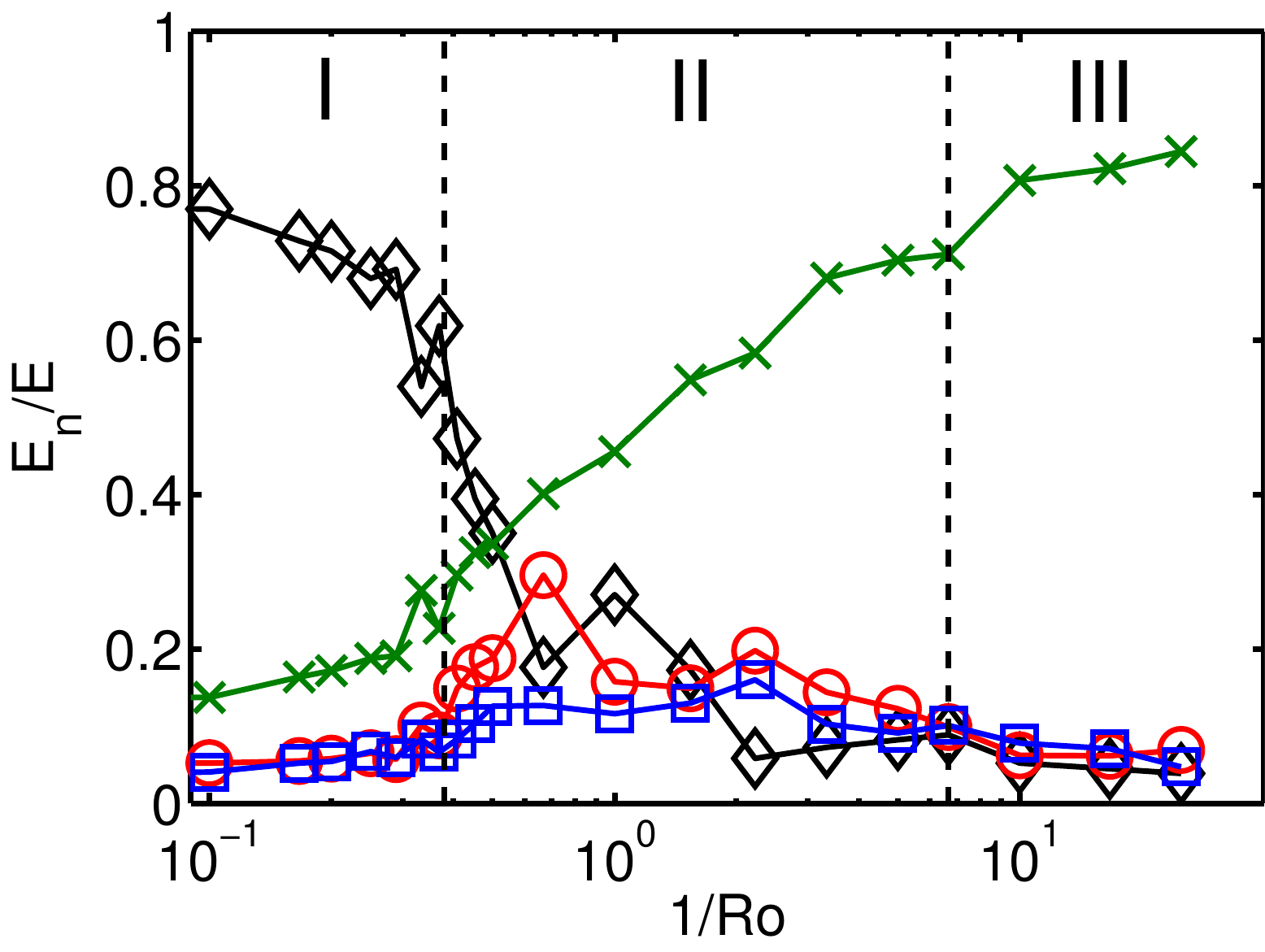}}
  \subfigure[]{\includegraphics[width=0.49\textwidth]{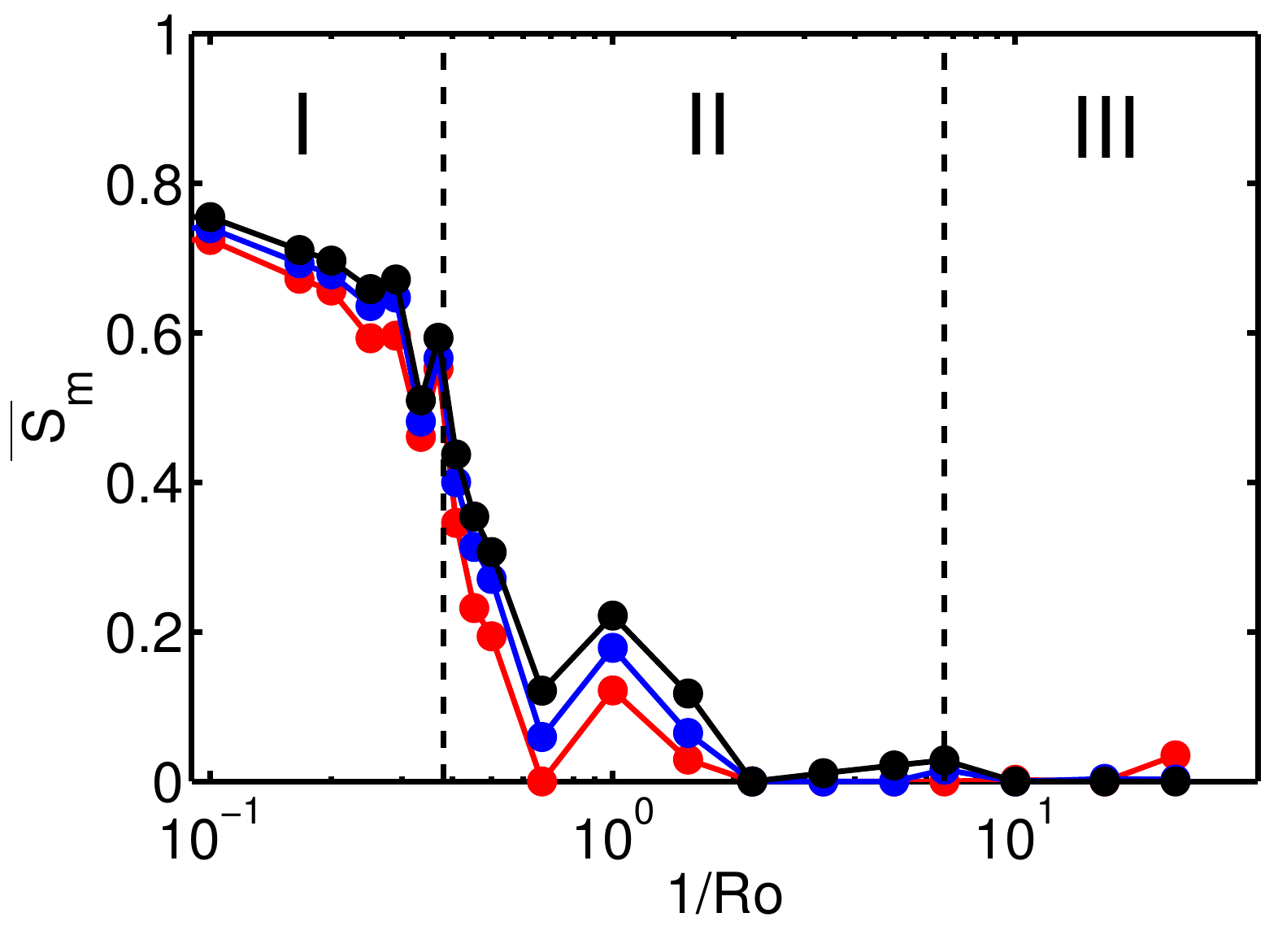}}
  \caption{Results from DNS on the magnitude of the different Fourier modes and 
$S_m$ for $Ra=2.73\times10^8$ and $Pr=6.26$ based on the data of the numerical 
probes at $z=0.50L$. Panels (a), (b), and (c) show the energy in the different Fourier 
modes of the azimuthal temperature profile normalized by the energy present in all 
Fourier modes based on the data of $8$, $16$, and $32$ equally spaced probes, 
respectively. The black, red, blue, and dark green lines indicate the energy in the 
first, second, third, and the additional modes, respectively. Note that when the data of 
$8$ probes are used a lot of information about the higher modes is lost. Panel d shows 
the corresponding relative LSC strength based on the data of $8$ (red), $16$ (blue), and 
$32$ (black) equally spaced probes. The two dashed vertical lines indicate the 
transitions between regimes I and II, and between II and III, respectively.}
  \label{Figure Fourier numerical}
\end{figure}

Figure \ref{Figure Fourier numerical} shows the magnitude of the different Fourier modes 
of the azimuthal vertical-velocity profiles for $Ra=2.73\times10^8$ and $Pr=6.26$ based 
on the data of $8$, $16$, and $32$ equally spaced probes. The result in panel (a) agrees 
well with the result obtained by \citet{kun08b}. Furthermore, a comparison of the relative 
energy in the different Fourier modes based on the data of $8$, $16$, and $32$ probes, 
see figure \ref{Figure Fourier numerical}, reveals that $8$ probes are insufficient to 
capture all flow characteristics. Since the results based on the data of $16$ and $32$ 
probes are very similar, we assume that $16$ probes should be sufficient to capture all 
relevant features of the azimuthal profiles.

Fortunately, the data of $8$ equally spaced thermistors are already sufficient to 
reliably calculate $S_m$ for non-rotating RB convection, see \citet{ste10c}. 
To confirm this observation for the RRB case (and validate the experimental 
results with $8$ probes discussed in section \ref{section Experiments}) we therefore 
calculate $S_m$ based on the data of $8$, $16$, and $32$ equally spaced probes. 
The result is given in figure \ref{Figure Fourier numerical}(d), which shows 
that the curves for the relative LSC strength almost collapse for the three cases. 
Moreover, the numerical data are in good agreement with the experimental result shown in 
figure \ref{Figure Fourier experimental}, including the large decrease of $S_m$ 
at the transition between regimes I and II. The small value of the relative LSC strength 
($S_m\ll 0.5$) suggests the absence of the LSC and that the azimuthal temperature profile at midheight becomes random in regime II and III. This is confirmed by a three-dimensional visualization of the flow, see figure 5 of \citet{zho09b}, where it is shown 
that in regime II vertically aligned vortices are the dominant feature of the flow. We note that figure~\ref{Figure Fourier numerical} shows a continuous decrease of the relative LSC strength, which is in disagreement with the measurements of~\citet{zho10c} that show a small increase in the temperature amplitude of the LSC between~$1/Ro=0$ and~$1/Ro\approx 0.41$. This difference might be caused by the slightly different monitoring locations (inside the sidewall in the experiments and at the radial position $0.45 L$ in the numerical simulations).

\section{The role of Ekman and Stewartson boundary layers} \label{section_Ekman_Stewartson}
\noindent In this section we will first discuss the flow characteristics for regimes I 
and II and sketch the influence of background rotation on the mean flow in general terms, 
and the role of the Ekman and Stewartson boundary layers is emphasized. Subsequently, we 
provide physical explanations for the structure of the mean flow as observed for regime 
II. Finally, we discuss the implications for the mean temperature gradient at the 
sidewall of the convection cell.

\subsection{Description of the flow characteristics}
\begin{figure}
  \centering
 \includegraphics[width =0.99\textwidth]{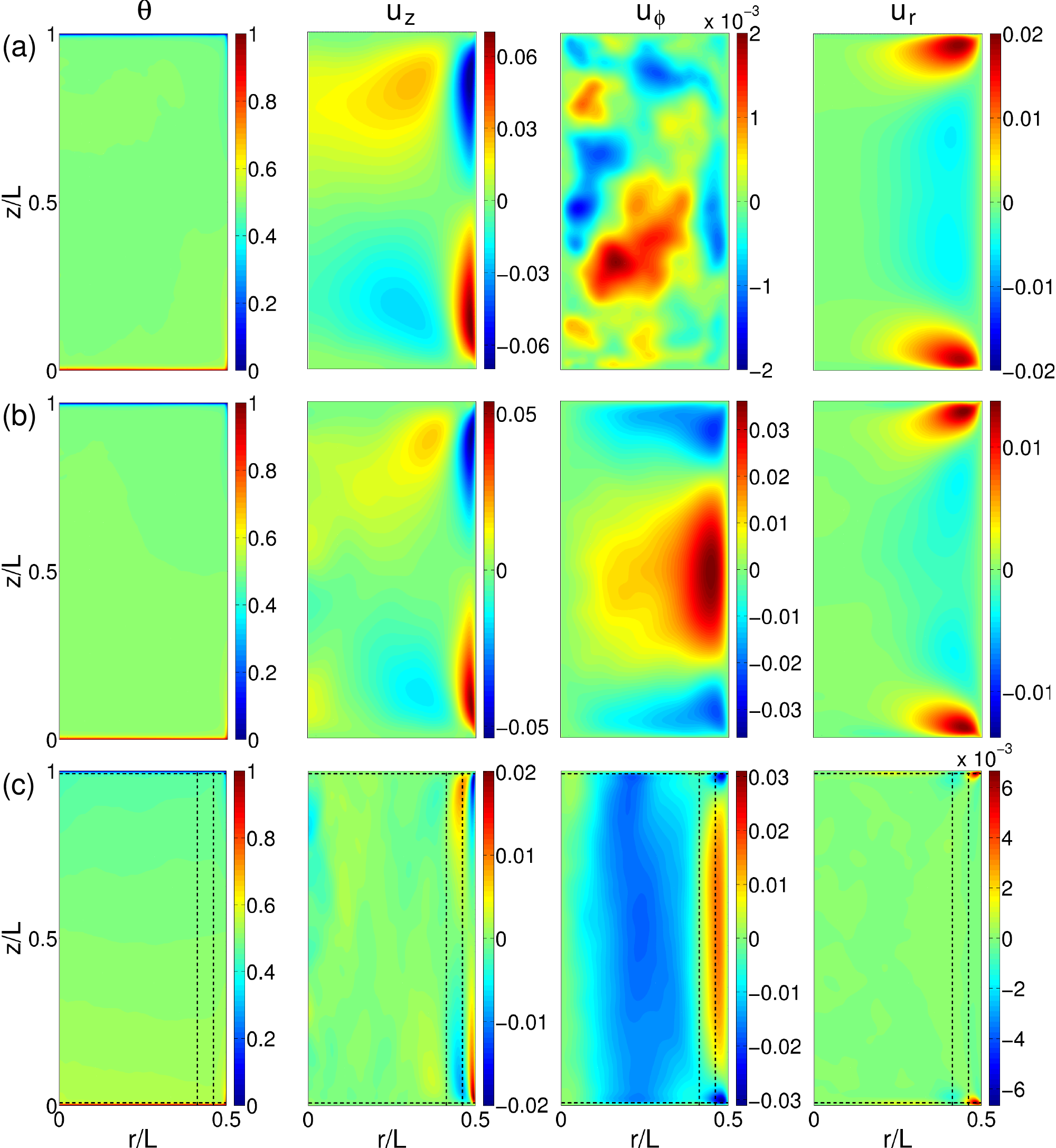}
  \caption{The figures from left to right indicate temporal and azimuthal averages of temperature and the velocity components $u_z$, $u_{\phi}$, and $u_r$, respectively, for $Ra=1.00 \times 10^9$ and $Pr=6.4$. The rows from top to bottom indicate the results for (a) $1/Ro = 0$, (b) $1/Ro = 0.35$, and (c) $1/Ro = 2.78$, respectively. The left-hand side of each picture is the cylinder axis $r = 0$; the right-hand side corresponds to the sidewall $r/L = 0.5$. The dashed lines in the pictures of the bottom row indicate typical boundary-layer thicknesses: $\delta_E$ = $Ek^{1/2}$ near the bottom and top plates, and $\delta_{S,1/3}$ = $Ek^{1/3}$ and $\delta_{S,1/4}$ = $Ek^{1/4}$ near the sidewall ($\delta_{S,1/3}$ is closest to the sidewall). Note that the azimuthally averaged azimuthal velocity for $1/Ro = 0$ (a) is much smaller than for $1/Ro = 0.35$ (b) and $1/Ro = 2.78$ (c).
  }
  \label{Figure averaged velocity}
\end{figure}

\begin{figure}
  \includegraphics[width=0.965\textwidth]{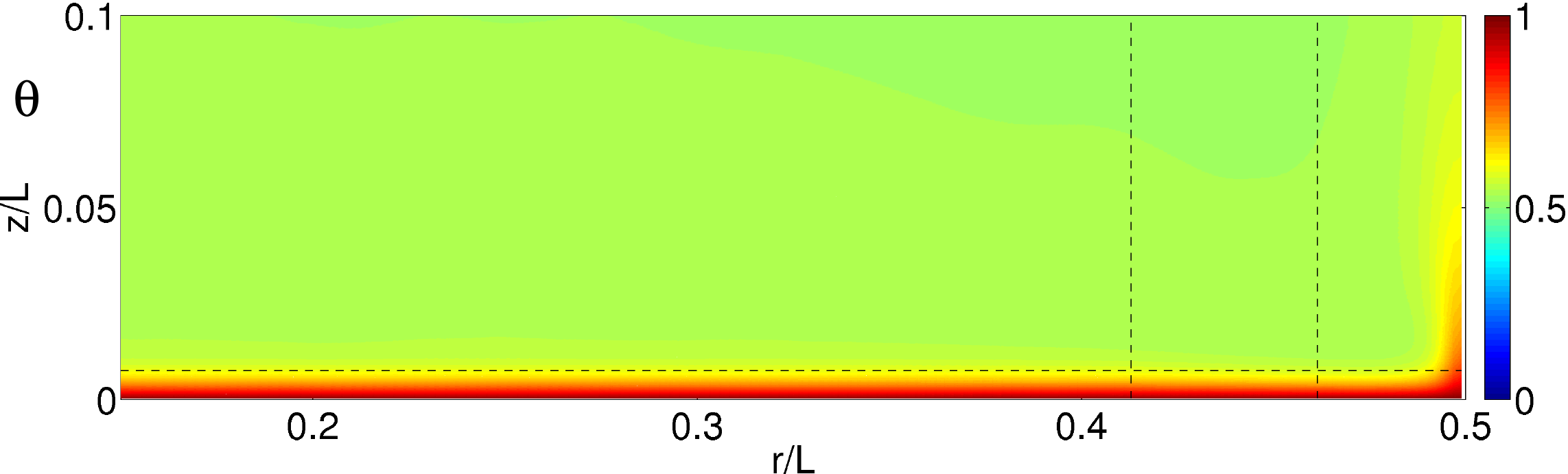}
  \includegraphics[width=0.99\textwidth]{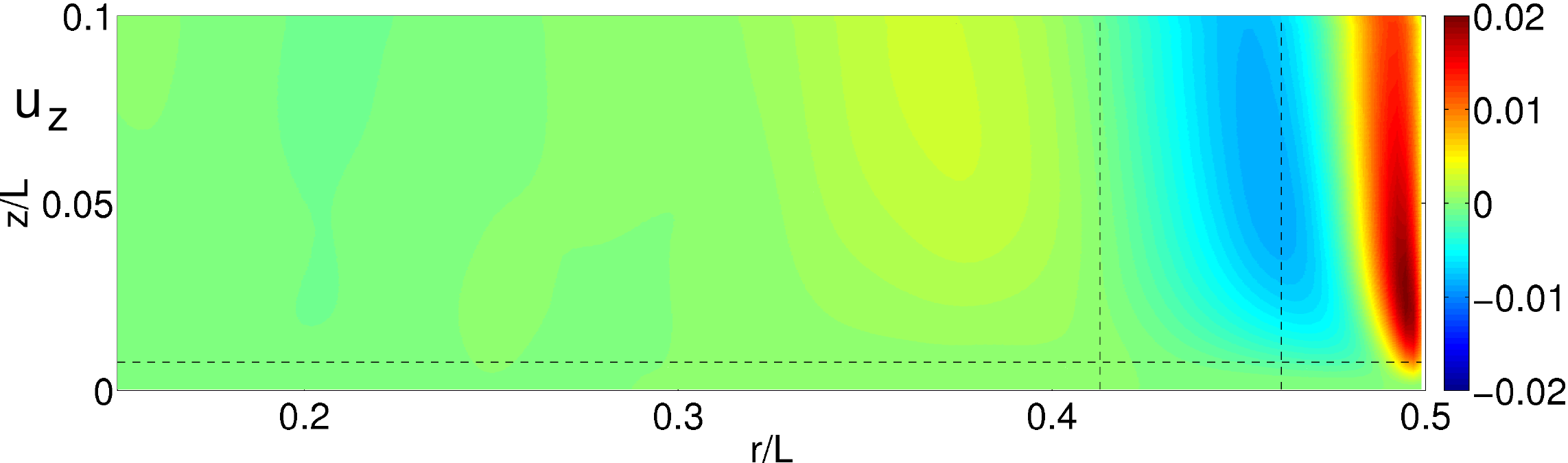}
  \includegraphics[width=0.99\textwidth]{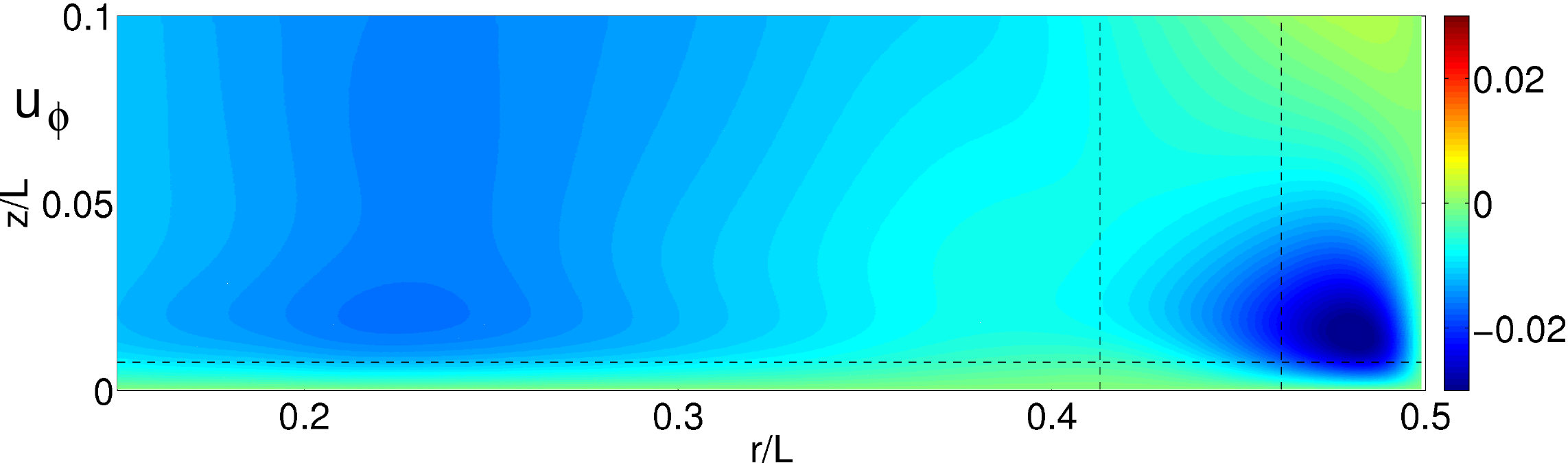}
  \includegraphics[width=0.99\textwidth]{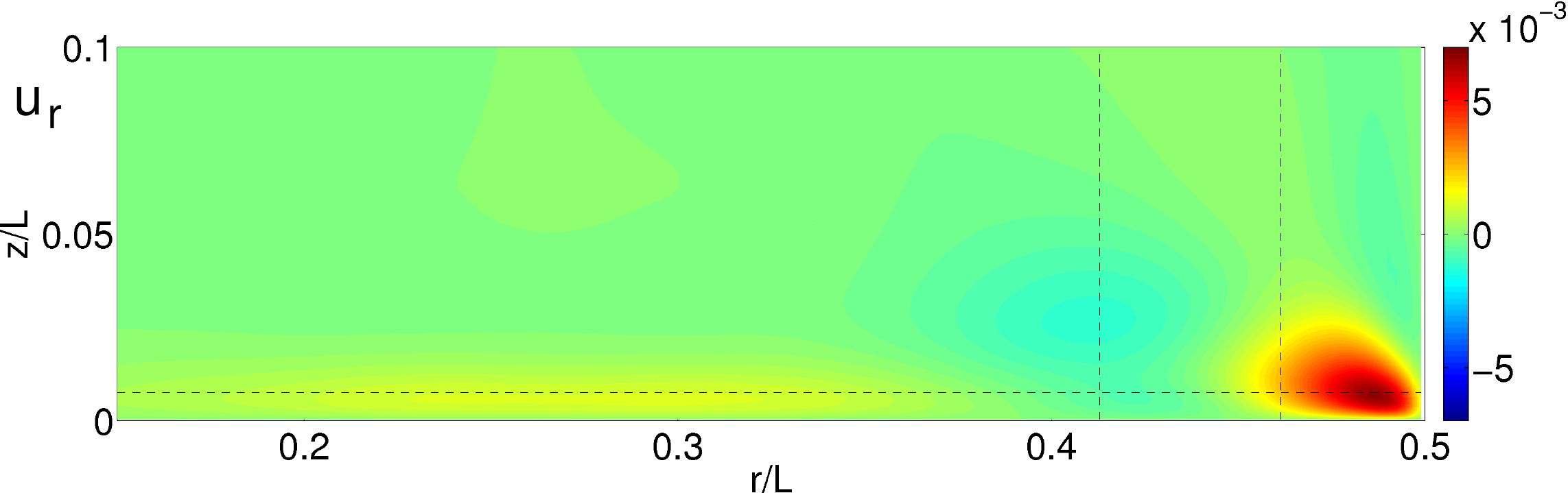}
\caption{Close-up of the bottom-right corner of the figures shown in figure \ref{Figure averaged velocity}(c) ($1/Ro = 2.78$). The area displayed is $0.15 < r/L < 0.5$; $0 < z/L < 0.1$. Color coding and dashed lines as in figure \ref{Figure averaged velocity}(c). The panels from top to bottom indicate the temporal and azimuthal averages of temperature and the velocity components $u_z$, $u_{\phi}$, and $u_r$, respectively.}
  \label{Figure averaged velocity zoom}
\end{figure}

In figure \ref{Figure averaged velocity} we show the azimuthally averaged temperature and 
velocity components $u_z$, $u_{\phi}$ and $u_r$ for three typical rotation regimes: 
(the top row) $1/Ro = 0$ (no rotation, regime I), (the middle row) $1/Ro = 0.35$ (weak rotation, regime I) and 
(the bottom row) $1/Ro = 2.78$ (moderate rotation, regime II) at $Ra = 1\times 10^9$ and $Pr = 6.4$. 
The left-hand side of each picture ($r = 0$) represents the position of the cylinder axis 
(dash-dotted line), while the right-hand side ($r/L = 0.5$) corresponds with the 
sidewall. In addition to the azimuthal averaginging, they have also been averaged in time for more than one hundred large-eddy turnover times to find the mean circulation hidden under the turbulent fluctuations. A close-up of the thermal and flow structure in the bottom corner of the tank 
for the case $1/Ro = 2.78$ (see figure \ref{Figure averaged velocity}c) is shown in figure \ref{Figure averaged velocity zoom}. We note that the 
averaged profiles for lower $Ra$ values are similar to the ones presented in 
figure \ref{Figure averaged velocity}.

\begin{figure}
  \centering
  \includegraphics[width=0.40\textwidth]{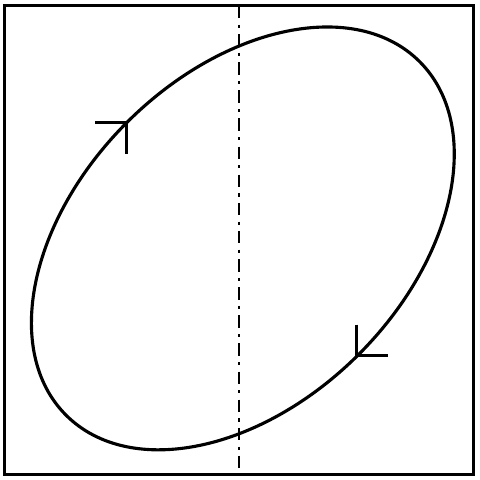}
  \caption{Schematic side view of the tilted LSC in the cylinder in the non-rotating or 
weakly rotating (regime I) case. The dash-dotted line is the axis of the cylinder.}
  \label{Figure LSCorientation}
\end{figure}

For the non-rotating case ($1/Ro = 0$), shown in \ref{Figure averaged velocity}(a),
we observe the signature of the LSC. When looking at a vertical cross-section of the domain, the cross-sectional plane aligned with the LSC, the mean circulation has an 
elliptic shape with its major axis oriented at some angle with the cylinder axis (see 
the sketch in figure \ref{Figure LSCorientation}, and also in \citet{qiu01b}). It is due 
to this mean tilt of the elliptic LSC, combined with the rotation sense as shown in 
figure \ref{Figure LSCorientation}, that the \emph{azimuthal} average of this flow is not zero; 
especially upward and downward motions are separated in  these azimuthally averaged 
plots. In the top half upward motion (positive $u_z$) is found near the cylinder axis, 
while the downward flow (negative $u_z$) is strongest near the sidewall. The opposite 
situation is found for $u_z$ in the bottom half. The azimuthal velocity $u_{\phi}$ does 
not show a well-defined mean profile, just some small fluctuations. The averaged radial 
velocity $u_r$ has maximal values near the intersections of the horizontal plates with 
the sidewall, consistent with the averaged vertical velocity $u_z$ near these regions. 
Note also the weak radially inward flow near the sidewall around midheight.

When a small rotation is introduced [$1/Ro = 0.35$, figure \ref{Figure averaged velocity}(b)], a first observation is that the mean velocities become smaller (except the averaged azimuthal velocity $u_{\phi}$ which grows due to organization of the azimuthal flow). The vertical velocity $u_z$ still has the dominant structuring due to the LSC, but especially near the cylinder axis some disturbances appear. For the azimuthal velocity a well-defined structure emerges due to the azimuthal averaging. The Coriolis force, deviating the parts of the LSC with mean horizontal velocity to the right with respect to the local direction of the (mean) horizontal flow, is driving the organization of an average azimuthal flow. The blue regions near the corners in the plot of the azimuthal velocity $u_{\phi}$ represent the anticyclonic motion induced by the mean horizontal outward flow of the LSC (near the bottom and top plates, see figure \ref{Figure LSCorientation}). The sketch of the LSC motion shows that the up (down) going plumes are first traveling straight up (down), before the radial inward flow of the LSC sets in. The cyclonic motion that is observed in the central sidewall region of the azimuthally averaged azimuthal velocity is due to the radial inward flow of the LSC in the top (bottom) half of the cell for the upward (downward) branch of the LSC, see figure \ref{Figure LSCorientation}. This radial inward flow of the LSC results in a spin-up, i.e. in cyclonic azimuthal motion, by the action of the Coriolis force. This mechanism is equivalent to conservation of  angular momentum: radial inward motion ($u_r<0$) of fluid parcels results in an increase of $u_{\phi}$, i.e. in cyclonic motion. Because this radial inward flow is in the top (bottom) part of the sample for the up (down) going plumes this results, after azimuthal averaging, in one central sidewall region in which a cyclonic motion is found. This means that this particular snapshot, cannot be used for interpretation of the movement of individual plumes traveling from bottom to top and the other way around. However, in this very schematic picture it does seem to be consistent with experimental measurements of the anticyclonic motion of the LSC \citep{har02,bro06b,kun08b,zho10c} as both upgoing and downgoing plumes get anticyclonic deflection by the mean anticyclonic azimuthal flow near the plates and any effect of mean cyclonic flow becomes effective only for $z\gtrsim2/3H$ (for upgoing plumes) and $z\lesssim H/3$ (for downgoing plumes).

At much higher rotation rates, i.e. for larger $1/Ro$-values, the LSC no longer exists, 
and the secondary circulation takes on a different appearance. It is obvious from the 
velocity plots that relatively large mean velocities occur in thin regions near the solid 
container walls. A division of the domain into several regions is appropriate, being the 
bulk interior domain, the Ekman layers at the horizontal plates, and the Stewartson layer 
at the sidewall. The Ekman layers have a (non-dimensional) thickness $\delta_E = Ek^{1/2}$ (with the Ekman 
number defined as~$Ek=\nu/(\Omega L^2)$, representing an inverse Reynolds 
number based on the length scale $L$ and velocity scale $L\Omega$). The Stewartson layer 
at the sidewall has a sandwich structure consisting of a thicker outer layer of non-dimensional thickness 
$\delta_{S,1/4} = Ek^{1/4}$  and a thinner inner layer of non-dimensional thickness 
$\delta_{S,1/3} = Ek^{1/3}$ \citep[see e.g.][]{ste57,ste66,moo69,hei83,hei84,hei86}. The dashed 
lines in figure \ref{Figure averaged velocity}(c) and figure \ref{Figure averaged velocity zoom} indicate these typical layer thicknesses. In figure 
\ref{Figure averaged velocity zoom} the bottom right-hand corner is magnified for the 
temperature (top panel) and each of the velocity components, so that the various boundary 
layers are more easily recognized. Note that since the secondary circulation associated 
with the Ekman and Stewartson layers is weak, the mean velocities $u_r$ and $u_z$ in the 
bulk are very small and therefore hardly noticeable, as is confirmed by the
simulation results. It should be emphasized that the secondary circulation found here 
is considerably different from that found by \citet{har99} or that of \citet{hom69}, both driven by
centrifugal buoyancy. In both these works the up-down symmetry is broken due to centrifugal accelerations having opposite directions near the plates, viz. cold fluid near the top plate accelerates radially outward while hot fluid near the bottom plate goes radially inward. Here we disregard centrifugal acceleration and up-down symmetry is preserved.
It must also be stated that the aforementioned circulation is only of secondary magnitude. The 
still-turbulent field of the vortical plumes is dominant.

The most remarkable change is observed in the distribution of the azimuthal velocity 
$u_{\phi}$: in the bulk of the domain an anticyclonic flow ($u_{\phi}<0$, indicated by 
blue in figure \ref{Figure averaged velocity}c) is present, while a thin band of cyclonic flow ($u_{\phi}>0$, 
indicated by red) is visible only close to the sidewall. Simulations for the parameter 
range $0.35 \lesssim 1/Ro \lesssim 2.78$ (not shown here) revealed that for increasing 
rotation rates (increasing $1/Ro$-values) the cyclonic flow region becomes confined to a 
region near the sidewall of increasingly smaller size. These observations suggest that 
in this $Ro$-regime the Ekman and Stewartson layers play an essential role in the mean 
secondary circulation, even though this circulation is relatively weak and even 
continuously disturbed by the non-steady thermally driven turbulent plumes that are 
present throughout the flow domain.

\subsection{Analysis secondary circulation regime II}
\begin{figure}
  \centering
  \subfigure[]{\includegraphics[width=0.49\textwidth]{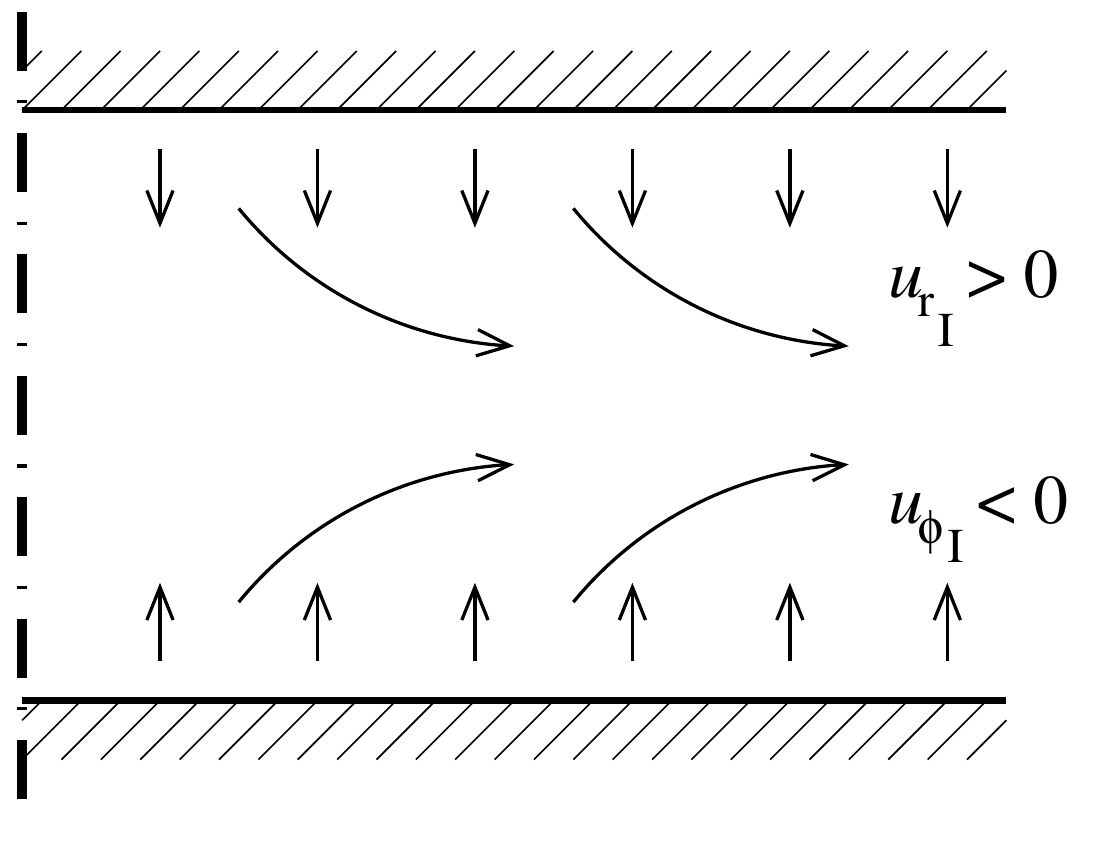}}
  \subfigure[]{\includegraphics[width=0.49\textwidth]{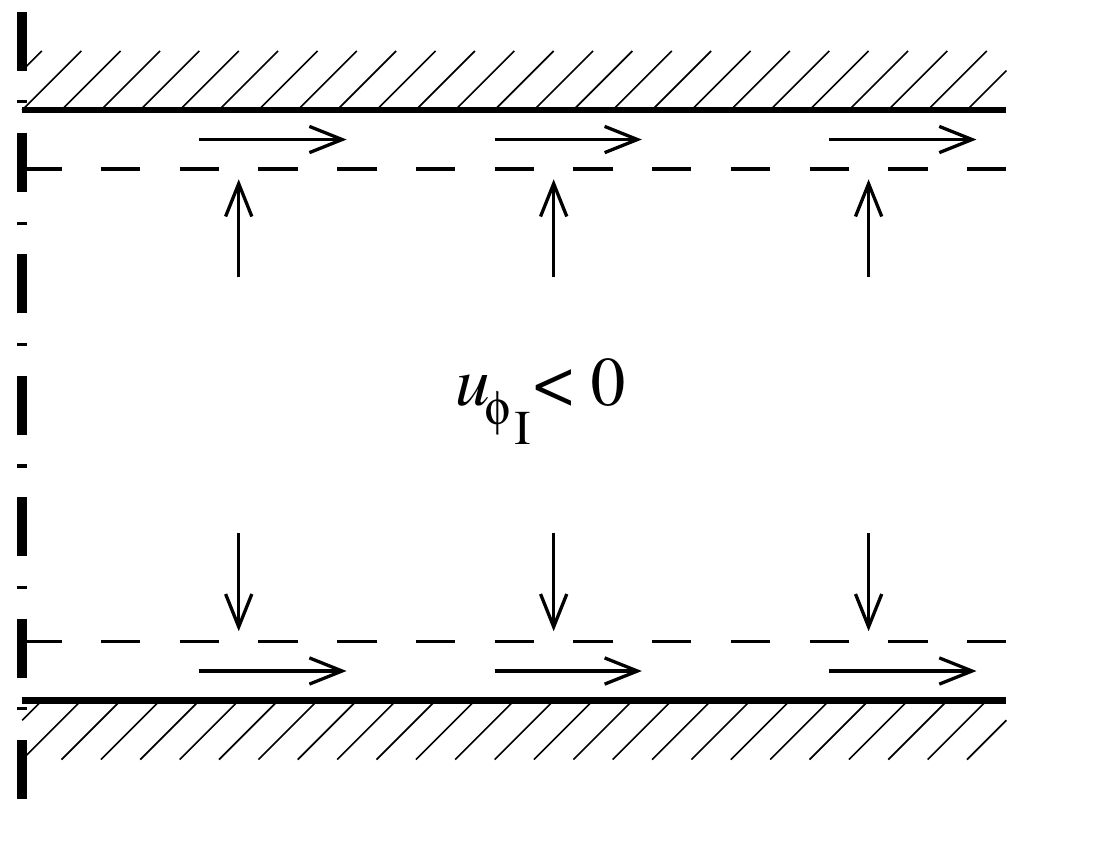}}
  \caption{(a) Fluid motion due to thermal forcing. The heated (cooled) fluid rising upward 
(flowing downwards) from the bottom (top) plate result in a radially outward mean flow in the bulk 
of the domain. (b) Fluid motion due to the suction of the Ekman layers (spin up). The Ekman layers impose 
suction towards the plates. Mass conservation implies a radial outflow inside these layers. The dash-dotted line in both panels indicates the center of the cell (which coincides with the rotation axis).}
  \label{fig:sketches}
\end{figure}

The feature of the weak anticyclonic swirl flow in the bulk of the domain in 
combination with a region of cyclonic swirling motion near the cylinder wall (the 
$u_{\phi}$-plot in figure \ref{Figure averaged velocity}(c)) is remarkable. Although the exact mechanism causing this bulk motion is an open question, we offer a possible explanation as follows. The heated fluid rising 
upward from the bottom plate and the cooled fluid flowing downward from the top plate 
result in a radially outward mean flow in the bulk of the domain, as sketched in the~$(r,z)$ plane in figure~\ref{fig:sketches}(a). Although this radial motion is relatively weak, conservation of angular momentum ($\sim rV$, with $V = \Omega r + u_{\phi}$ the absolute azimuthal velocity) implies that the absolute 
swirl velocity of fluid parcels will decrease, i.e. in the co-rotating frame this fluid 
will acquire an anticyclonic azimuthal motion ($u_{\phi}<0$). This radial outflow in the 
bulk of the domain is of such small magnitude that it leaves no strong trace in the azimuthally averaged radial velocity~$u_r$. Furthermore, the plots of averaged vertical velocity~$u_z$ show no mean motion directed away from the horizontal plates: this transport is localised in the small thermal vortices that vanish after azimuthal averaging.

This picture of the flow is not complete, however, because a second mechanism is active 
simultaneously in the bulk of the domain: the spin-up process~\citep{gre63}. This second process is 
driven by the Ekman layers (figure~\ref{fig:sketches}b): the bottom Ekman layer imposes a suction 
velocity on the interior given by $u_z=\frac{1}{2}Ek\,\omega_I$ in dimensionless form, 
with $\omega_I$ the vertical component of the rotation of the relative interior flow. To clarify, as~$\omega_I$ is anticyclonic, the corresponding vertical velocity according to the suction condition is directed from the bulk into the Ekman layer. A 
similar suction velocity is imposed on the interior flow by the upper Ekman layer, also directed from the bulk to the boundary layer. There is thus a mean flow from the bulk into the Ekman layers. By conservation of mass, a radial outflow in the Ekman layers is needed to carry away the flux from the bulk into the Ekman layers. The radial outflow is easily identified in the averaged radial velocity plots due to the tiny radial cross-sectional area of the Ekman layer, leading to a relatively large mean velocity.

The radially outward transport through the Ekman layers is returned via a Stewartson boundary layer at the cylinder wall. This layer has a small mean radial inward transport into the interior distributed along its vertical extent. An indication of this radially inward transport is found in the~$u_r$ plot of figure~\ref{Figure averaged velocity zoom} at~$(r,z)\approx(0.4, 0.025)$. There is also an internal recirculation in this layer: Again, according 
to the principle of conservation of angular momentum $rV$, with $V = \Omega r + u_{\phi}$, the 
radial inward motion from the sidewall boundary layer in the bulk implies a change in the swirl velocity, which is reduced in magnitude and approaches zero.

It can be shown that the matching of the azimuthal velocity component $u_{\phi}(r = 0.5L)$ to the sidewall 
requires a Stewartson layer of typical thickness~$Ek^{1/4}$, in which this velocity is given by 
$u_{\phi} = -\frac{1}{L} u_{\phi} (r=0.5L) \exp ( \xi \sqrt {2/L} )$, with $L$ the 
height of the cylinder, and $\xi = (r - 0.5L) Ek^{-1/4}$ the stretched radial 
boundary-layer coordinate~\citep{gre63}. When referring to this principal velocity component as being 
${\mathcal{O}}(1)$, the vertical and radial velocity components in this layer are much 
smaller, viz. ${\mathcal{O}}(Ek^{1/4})$ and ${\mathcal{O}}(Ek^{1/2})$, respectively (in 
non-dimensional terms). The radial velocity is also correctly matched to the sidewall. However,
matching of the vertical velocity to the cylinder wall requires the 
presence of a thinner $Ek^{1/3}$ layer \emph{inside} the $Ek^{1/4}$ layer. The vertical matching implies a principal solution
with a vertical velocity $u_z \sim {\mathcal{O}}(Ek^{1/4})$ within this layer. However, the entire velocity field in the~$Ek^{1/3}$ 
layer carries no net vertical flux; it is just an internal recirculation~\citep{gre63}.

We will now describe the internal recirculation in more detail. Where the Stewartson $Ek^{1/3}$ layer
meets the Ekman layers, the vertical velocity has 
a singular structure~\citep{moo69,hei83,hei86}: $u_z \sim \pm \delta(\eta)$ at $z = 0$ and $L$, with 
$\eta = (r - 0.5L)Ek^{-1/3}$ the stretched radial coordinate in this Stewartson layer, 
and $\delta$ the Dirac delta function. Signs of these singular eruptions of the Ekman 
layer fluxes at $z = 0$ and $z = L$ into the thinner Stewartson layer on the sidewall are clearly visible 
in the $u_z$-plot in figure \ref{Figure averaged velocity}(c), and for the case of the 
bottom corner in figure \ref{Figure averaged velocity zoom}. Note that these vertical 
fluxes are positive and negative near the bottom and top Ekman layers, respectively. 
Weaker vertical velocities of opposite signs are observed in the thicker 
$Ek^{1/4}$ layer, see figure \ref{Figure averaged velocity}(c). 
This oppositely signed vertical velocity next to the peak corresponds with the internal recirculation of the singular upward flux. 
Furthermore, the radial branches of the internal recirculation can be recognised in the~$u_\theta$ plots. Close to the bottom and top plates there is strong radially outward flow in the thin Ekman boundary layers, which corresponds to strongly negative azimuthal velocity (corner regions near the sidewall of figures~\ref{Figure averaged velocity}(c) and~\ref{Figure averaged velocity zoom}). Along the sidewall away from the plates there is radial inflow to close this recirculation, which, due to the much larger vertical extent, is of such small magnitude that it is not found in the plot of mean radial velocity. However, the radial inflow leaves its signature on the mean azimuthal velocity: radial inflow corresponds to positive~$u_\theta$, which is observed as the orange band near the sidewall in the~$u_\theta$ plot of figure~\ref{Figure averaged velocity}(c). 

\begin{figure}
  \centering
  \includegraphics[width=0.5\textwidth]{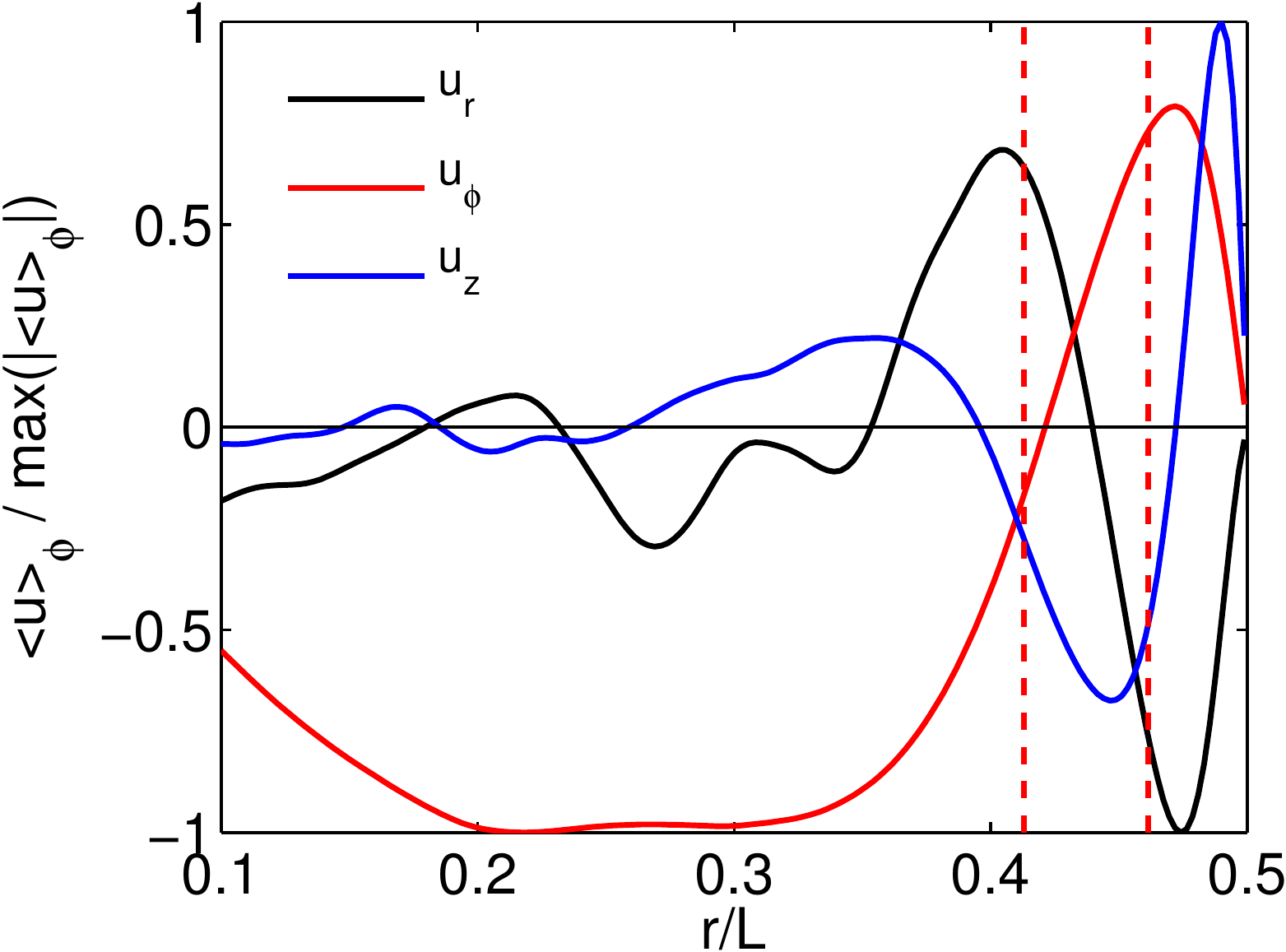}
  \caption{The azimuthally averaged velocity profiles at $z=0.25L$ for $Ra = 1.00 \times 10^9$ and
$Pr = 6.4$ for $1/Ro = 2.78$. The azimuthal, radial, and axial velocity components are indicated in red, black, and blue, respectively.}
  \label{fig:Rossby0.30}
\end{figure}

To further illustrate the secondary circulation we consider radial mean profiles of velocity. Figure \ref{fig:Rossby0.30}(a) shows the azimuthally averaged velocity profiles $u_r(r)$, 
$u_{\phi}(r)$, and $u_z(r)$ at $z = 0.25L$ according to a numerical simulation for 
$1/Ro = 2.78$. The radial $u_{\phi}$-distribution clearly shows the presence of 
anticyclonic swirl in the interior domain, while significant positive swirl exists in 
the region close to the cylinder wall. The vertical velocity shows the peaked structure close to the cylinder wall, indicating the singular eruption of the Ekman flux into the $Ek^{1/3}$ layer, with accompanying negative velocity directly adjacent to close the recirculation. Furthermore, the radial inflow in the sidewall region can be observed here, corresponding to positive~$u_\theta$. In this normalised plot this radial-velocity structure may be recognised, yet in absolute sense the radial velocity component is much smaller than the vertical and azimuthal contributions.

The description of the secondary circulation is only qualitative, as a full quantitative treatment would lead too far for the purpose of the current paper. A quantitative treatment of the secondary circulation will be presented in a forthcoming paper.

\begin{figure}
  \centering
  \includegraphics[width=0.49\textwidth]{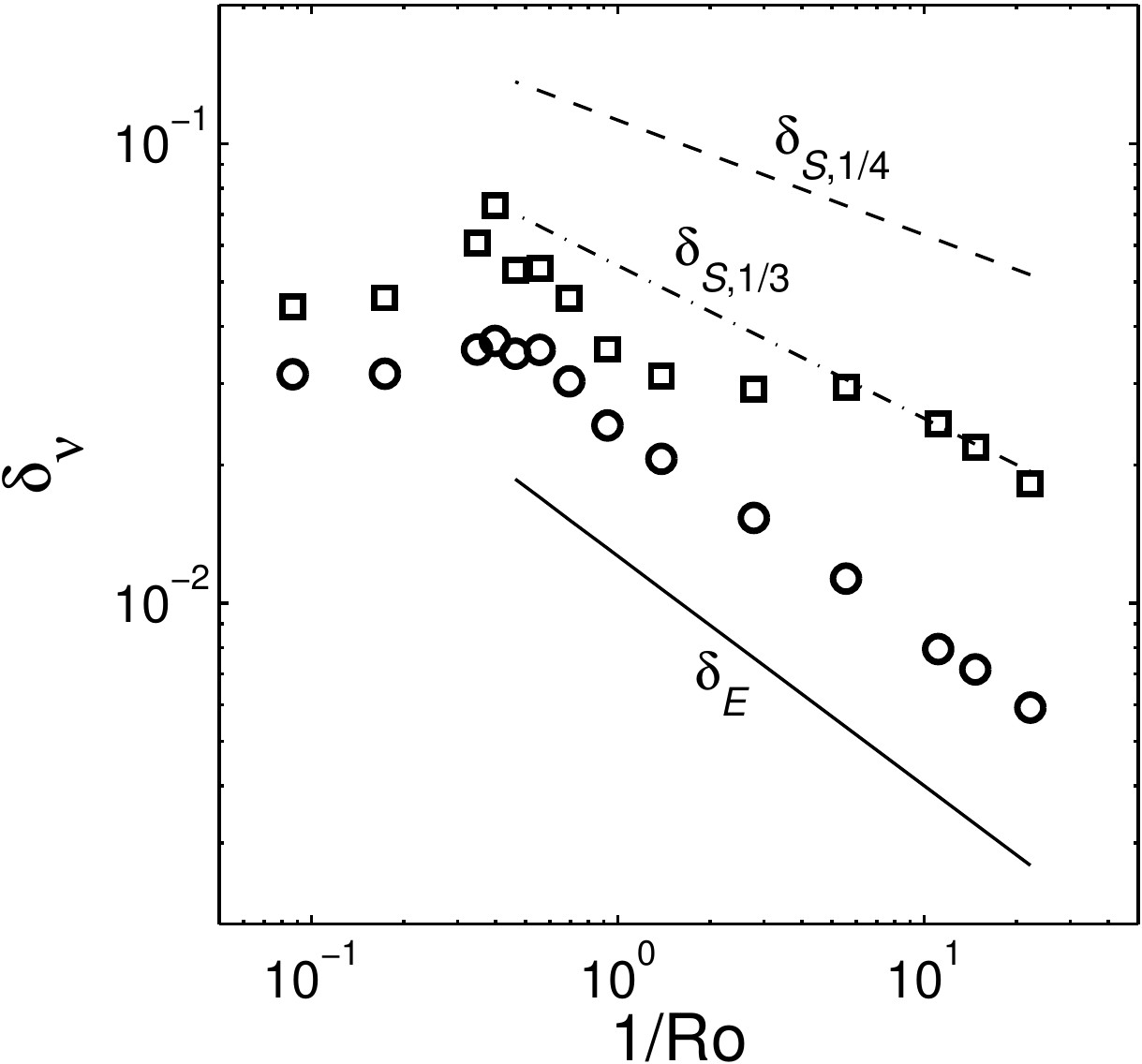}
  \caption{Boundary layer thicknesses for $Ra=1.00\times10^9$ and $Pr=6.4$. The 
dimensionless viscous boundary layer thickness near the plates is indicated by the 
circles. The dimensionless boundary layer thickness of the viscous boundary layers near 
the sidewall, i.e. the Stewartson layers, is indicated by the squares and follows the 
scaling of the inner Stewartson boundary layer thickness, i.e. $\delta_{S,1/3}$. Adapted 
from \citet{kun10}.}
  \label{fig:Stewartson}
\end{figure}

Additional evidence that the mean velocity profiles close to 
the sidewall are indeed governed by Stewartson boundary layer dynamics is provided in 
figure \ref{fig:Stewartson}, where it is shown that the thickness of the sidewall 
boundary layers observed in the simulations follows the theoretical predictions~\citep{kun10}.

The role of Ekman and Stewartson boundary layer dynamics on the average mean flow is here illustrated
with numerical simulations with $Ra=1\times 10^9$ and $Pr=6.4$. Separate sets of simulations with
$Ra=2.73\times 10^8$ with $Pr=6.26$ and $Ra=1\times 10^8$ with $Pr=6.4$
essentially revealed the same picture. The explanation provided in this section is
therefore applicable to a range of Rayleigh numbers. Further studies are needed to explore the
regime with $Ra\gtrsim 1\times 10^9$ and the role of the Prandtl number. We have also 
explored the role of the Rossby number, in particular by increasing the value of $1/Ro$ 
such that the flow is dominated by strong rotation (close to and even in regime III). 
The vertical vortex plumes then become stronger and data from azimuthally averaged 
components of velocity reveal an increasing number of alternating, vertically almost homogeneous 
regions of $u_{\phi}<0$ and $u_{\phi}>0$.

\subsection{Influence of the secondary circulation on sidewall temperature measurements}
\begin{figure}
  \centering
  \subfigure[]{\includegraphics[width=0.49\textwidth]{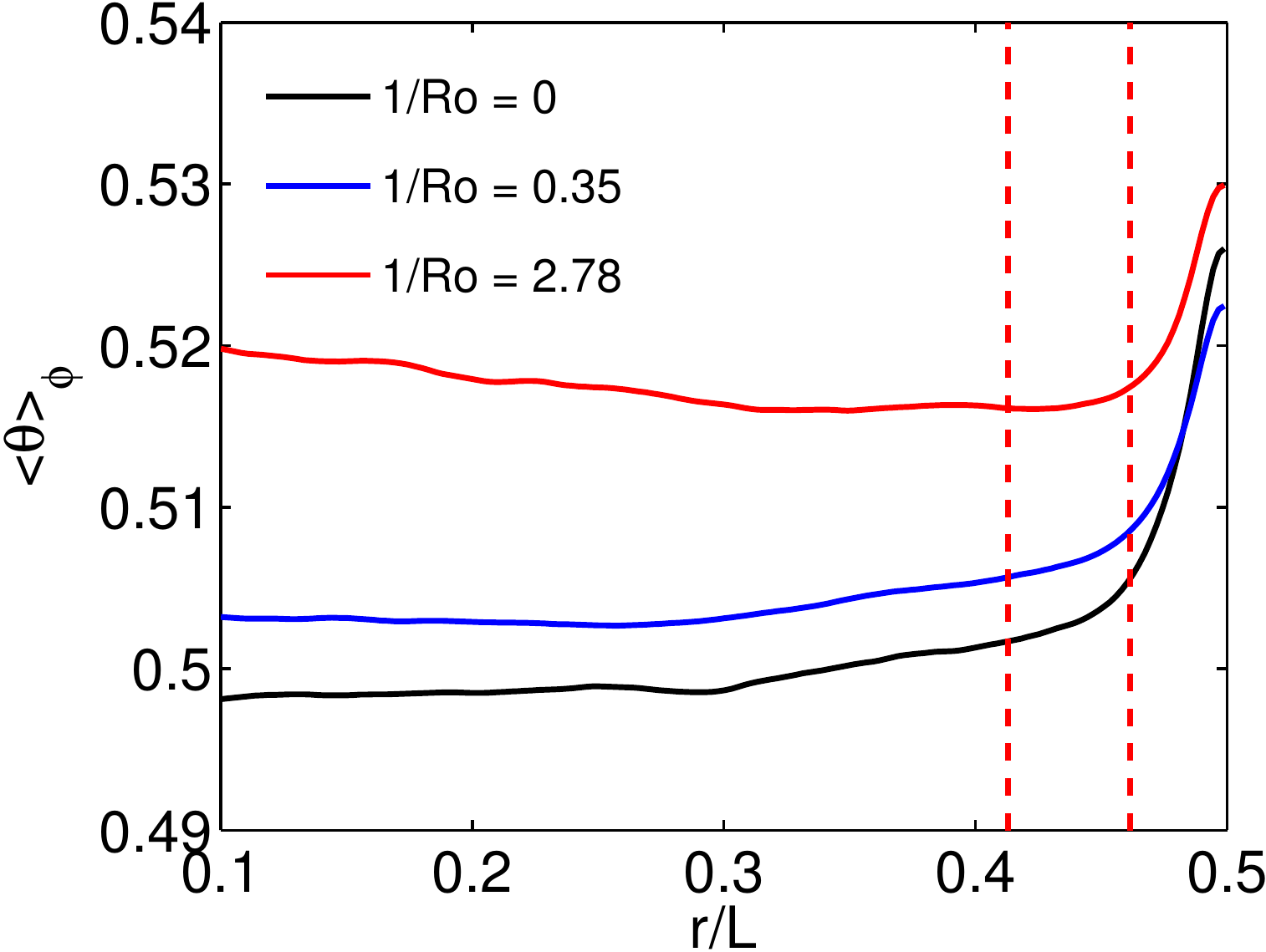}}
  \subfigure[]{\includegraphics[width=0.49\textwidth]{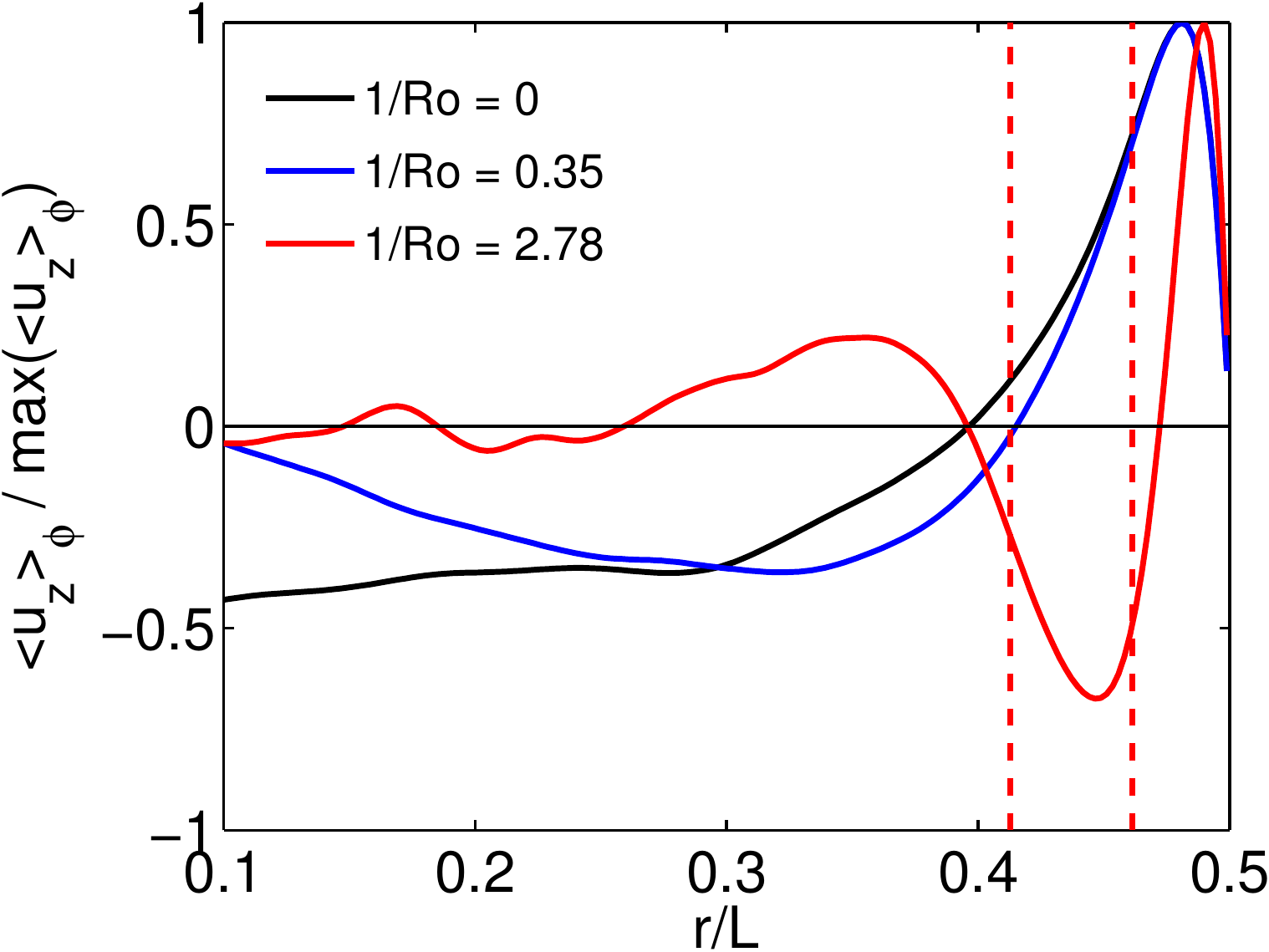}}
  \caption{(a) Azimutally averaged temperature and (b) vertical velocity profiles as 
function of the radial position at the height $z=0.25L$ for $Ra=1.00\times10^9$ and 
$Pr=6.4$. Black, blue, and red indicate the data for $1/Ro=0$, $1/Ro=0.35$, and $1/Ro=2.78$, 
respectively. The dashed lines indicate typical Stewartson boundary layer thicknesses 
($\delta_{S,1/3}$ = $Ek^{1/3}$ and $\delta_{S,1/4}$ = $Ek^{1/4}$ near the sidewall for 
$1/Ro=2.78$ ($\delta_{S,1/3}$ is closest to the sidewall). Note that for $1/Ro=2.78$ 
there is a strong upward flow of fluid in the inner Stewartson $Ek^{1/3}$ layer, which causes 
a vertical temperature gradient at the sidewall. In panel (a) the increasing mean temperature in the bulk with increasing rotation rate 
indicates that the vertical temperature gradient in the bulk increases with 
increasing $1/Ro$. The temperature gradient in the bulk for this case is shown in figure 15 of \citet{kun10}.
 }
  \label{Figure Radial tempprofile}
\end{figure}

The above observations of the flow structure enable us to explain the sidewall 
temperature measurements. In figure \ref{Figure Radial tempprofile} we have plotted 
radial distributions of the azimuthally averaged temperature and vertical velocity at 
$z = 0.25L$. Three main observations from the data extracted at this particular level 
are: a strong vertical mean flow near the sidewall (and for the highest rotation rate 
clearly within the $Ek^{1/3}$ layer), an increase of the mean temperature in the bulk 
with the rotation rate, and an increased mean temperature near the sidewall compared to 
the bulk, regardless of the existence of the LSC. We start with the latter observation. 
In regime I (weak rotation) where the LSC is the dominant feature of the flow, the 
temperature gradient at the sidewall is caused by the LSC, which carries warm fluid 
upwards along the sidewall and cold fluid down in the middle (when considered in an azimuthally-averaged plot); and vice versa in the top 
half of the cylindrical domain. In regime II (moderate rotation) the LSC has disappeared 
and vertical vortices form the dominant feature of the flow. In this regime the secondary 
circulation described above causes a flow directed vertically away from the plate close 
to the sidewall, i.e. the singular eruption into the $Ek^{1/3}$ layer, see figure \ref{Figure Radial tempprofile}(b). This means that in regime II the secondary 
circulation carries warm fluid upwards along the sidewall in the lower half of the 
cylinder and cold fluid downwards along the sidewall in the top half. The vertical temperature gradient on the sidewall is increased.
The enhanced mean temperature in the bulk with increasing rotation rate 
indicates that the vertical temperature gradient in the bulk increases also with 
increasing $1/Ro$, see for example figure 15 of \citet{kun10}. The bulk temperature gradient is caused by the merger of vertical 
plumes~\citep[see, e.g.,][]{jul96,jul99,leg01,spr06,bou86,zho93,eck98,spr06}, i.e. the enhanced 
horizontal mixing of the temperature anomaly of the plumes results in a mean temperature 
gradient. In regime III this temperature gradient in the bulk becomes stronger than in 
regime II and eventually becomes equal to the temperature gradient at the sidewall.

\section{Conclusions}
Based on the experimental data and results obtained from DNS 
we studied the characteristics of the azimuthal temperature profiles at the 
sidewall. We find that in regime I (weak rotation) the LSC is 
the dominant feature of the flow. In the sidewall temperature measurements this is 
identified by the strong presence of the first Fourier mode in the signal. When the 
rotation is increased a sudden onset in the heat transport enhancement is found that 
indicates the beginning of regime II, the moderate rotation regime. In this regime the 
LSC is replaced by vertically aligned vortices and due to their random position the magnitude of 
the first Fourier mode is very weak, which confirms that the LSC has disappeared. This 
feature is also observed in regime III, where the heat transfer decreases, because the 
vertical velocity fluctuations are suppressed by the rotation.

When the LSC is the dominant feature of the flow the vertical temperature gradient at the sidewall is mainly due to plumes that travel close to the sidewall because they travel with the LSC. However, also in regime II and III, in which the LSC is absent, there is a strong temperature gradient at the sidewall. In these regimes, where the vertical vortices are the dominant feature of the flow, the observed vertical temperature gradient along the sidewall is mainly due to the
secondary flow. The secondary flow, driven by the Ekman boundary layers near the plates, causes a recirculation in the Stewartson boundary layer on the sidewall with upward (downward) transport of hot (cold) fluid close to the sidewall in the bottom (top) part of the cell.

It is remarkable that the secondary flow in turbulent RB convection, which is observed after time-averaging of the flow field, is so well described by linear Ekman and Stewartson boundary layer theory. It seems that the mean laminar flow profiles prevail, hidden under highly turbulent fluctuations. Thus the knowledge about laminar flow dynamics can still be highly relevant in the study of turbulent flows. An example is the Grossmann--Lohse (GL) theory~\citep[see][for an overview]{ahl09}, in which the kinetic energy and thermal variance dissipation rates have been decomposed into boundary-layer and bulk contributions. Scaling-wise and in a time-averaged sense a laminar Prandtl--Blasius profile is assumed close to the horizontal plates. The GL theory successfully describes the~$Nu$ and~$Re$ number dependences on~$Ra$ and~$Pr$. Recently, \citet{zho10} and~\citet{zho10b} have shown that the laminar PB profile is indeed hidden under the turbulent fluctuations.

\acknowledgments

\section{Acknowledgements}
\noindent We gratefully acknowledge various discussions with Guenter Ahlers and Detlef Lohse over this line of research and their helpful comments on our manuscript. The authors wish to thank Eric de Cocq, Gerald Oerlemans, and Freek van Uittert (design and manufacturing of the experimental set-up) for their contributions 
to this project. Jaap van Wensveen of Tempcontrol is thanked for advice and for his help with the calibration of the 
thermistors. We thank Roberto Verzicco for providing the numerical code. R.J.A.M.S  and 
R.P.J.K. wish to thank the Foundation for Fundamental Research on Matter (Stichting voor 
Fundamenteel Onderzoek der Materie, FOM) for financial support. This work was sponsored 
by the National Computing Facilities Foundation (NCF) for the use of supercomputer 
facilities (Huygens), with financial support from the Netherlands Organization for 
Scientific Research (NWO).

\bibliographystyle{jfm}

\end{document}